\documentclass[aps,prb,twocolumn,superscriptaddress,showpacs,english,10pt]{revtex4-2}

\usepackage[T1]{fontenc}
\usepackage{babel}
\usepackage{amsmath}
\usepackage{amssymb}
\usepackage{wasysym}
\usepackage{graphicx}
\usepackage{xcolor}
\usepackage{braket}
\usepackage{multirow}
\usepackage{siunitx}
\usepackage[version=4]{mhchem}
\usepackage[inline]{enumitem}

\usepackage[linktocpage=true,
  colorlinks=true, 
  pdfborder={0 0 0},
  linkcolor=blue,
  citecolor=red,
  filecolor=yellow,
  urlcolor=blue,
  bookmarks,
  pdfauthor={},
]{hyperref}

\newcommand{\Graz}{Institute of Theoretical and Computational Physics, Graz University of Technology, NAWI Graz, 8010 Graz, Austria}

\begin{document}

\title{Superconductivity and strong anharmonicity in novel Nb-S phases}

\author{Roman Lucrezi}
\affiliation{\Graz}
\author{Christoph Heil} \email{christoph.heil@tugraz.at}
\affiliation{\Graz}

\date{January 11, 2021}

\begin{abstract}

In this work we explore the phase diagram of the binary Nb-S system from ambient pressures up to \SI{250}{\giga\pascal} using \textit{ab initio} evolutionary crystal structure prediction. We find several new stable compositions and phases, especially in the high-pressure regime, and investigate their electronic, vibrational, and superconducting properties. Our calculations show that all materials, besides the low-pressure phases of pure sulfur, are metals with low electron-phonon coupling strengths and critical superconducting temperatures below \SI{15}{\kelvin}. Furthermore, we investigate the effects of phonon anharmonicity on lattice dynamics, electron-phonon interactions, and superconductivity for the novel high-pressure phase of \ce{Nb2S}, demonstrating that the inclusion of anharmonicity stabilizes the lattice and enhances the electron-phonon interaction.
\end{abstract}

\maketitle

\section{Introduction}
In recent years, transition metal chalcogenides (TMC) have gained significant research interest based on their chemical and physical variety, as well as their tendency to create layered structures~\cite{tremel1995TMCreview,ivanova2019sesqui}. This constraint in geometry has been shown to be the source of many interesting phenomena related to electron-phonon ($ep$) processes such as charge density waves (CDW) and superconductivity (SC)~\cite{zettl1982NbS3CDW,Malliakas2013NbSe2CDW,shi2015SCTMD,gui2020Ta2Se}.
For some TMC materials, Mott-insulating behaviour or potential topological SC has also been predicted~\cite{Sipos2008TaS2Mott,wang2018TaS2,Li2018topoMoTe2}, properties that could find future applications in high-performance electronics, as controllable Mott transitions allow very fast, precise and efficient field-effect transistors (FET)~\cite{Zhou2013corrFETreview}, and Majorana states in topological SC could smooth the way to fault-tolerant quantum computing~\cite{Alicea_2012MajoranaTopoSC,Zhang2018TopoSC}.

Extensive studies have been conducted specifically on transition metal dichalcogenides (TMD). Featuring weakly bound van-der-Waals (vdW) layers, they offer a plethora of ways to manipulate and influence their physical properties, for example by varying the stacking order of the layers, intercalating other elements, doping via ionic liquids, etc.~\cite{klemm2015TMC2,Jung2016Intercalation,Wang2020NbS2}. This leads to applications as cathodes for batteries~\cite{Wittingham2004MoS2CathodeBattery}, solid lubricants due to low friction between the layers~\cite{Lee2010ThinMoS2Friction,Scharf2017TMDlubricant}, electrocatalysts for energy conversion~\cite{Voiry2016TMD2Dcatalyst,Xue2016TMDcatalyst}, and makes them versatile candidates for 2D beyond-graphene materials~\cite{Heine2015BeyondGrapheneTMD,Dai2016TMTCbeyondGrapheneTiS3,DRYFE2019BeyondGraphene2D}.

The semiconducting phases with group\nobreakdash-4 and group\nobreakdash-6 transition metals are studied for their applications in photovoltaics and electronics~\cite{Wang2012electroTMD,Wang2015tuningTMDs,zhou2017tetTMC}. Many TMDs undergo an indirect-to-direct band gap transition with decent carrier mobilities when bulk materials are exfoliated down to the monolayer limit~\cite{Mak2010MoS2DirectSemi,Ellis2011MoS2indirectDirectSwitch,Kumar2012direct1HMX2,Zhang2014DirectIndirectMoSe2Measurement}, making them favourable as FET components in terms of efficiency as well as size~\cite{Radisavljevic2011MoS2transistor,Ahmed2017TMDasFET,Chen2020TMDMosfetReview,Liu2020TMDcarriermobFET}.

Group\nobreakdash-5 TMDs, by contrast, can be metallic by filling the lowest $d$ band and exhibit conventional $ep$-mediated SC and CDW order. The coexistence and interaction of these two, at first glance mutually exclusive phases, is the focus of intensive research at the moment. Recent studies suggest that the $ep$ coupling responsible for creating the CDW phase is strongly localized in $\mathbf{k}$-space, allowing for the remaining $ep$ interactions at other wave vectors to promote SC~\cite{Valla2004NbSe2,Weber2011CDW,Wezel2011CDWTMD,tissen2013NbS2,liu2014NbS2,Liu2016TaSe2CDW,Ugeda2016NbSe2,heil2017NbS2,LianHeil2019CDWTaSe2}. 

External pressure can have significant effects on many physical properties, as it strongly influences atomic bonding and hence electronic properties, lattice dynamics, and $ep$ interactions, driving materials towards or away from lattice instabilities~\cite{lorenz2004HPSC,SuderowTissen2005pressureNbSe2,Calandra_CDW_2011,Leroux2015NbSe2pressure,Wang2017pressureTaSe2,Wang2017pressureWSe2,Ying_unusual_2018}.
While many TMD phases are well studied up to higher pressures, including investigations of their electronic and vibrational properties, and phase diagrams of other specific TMC stoichiometries at lower pressures have been reported~\cite{biberacher1980Nb3S4SC,zettl1982NbS3CDW,Dobashi2007Nb3S4SC1d,bloodgood2018NbS3,ivanova2019sesqui}, comprehensive and systematic studies on possible stable phases up to high pressures are still missing to the best of our knowledge.

With this work we want to contribute to fill this void by analysing the high-pressure phase diagram of Nb-S. We find several novel materials and phases, for which we determine their thermodynamic stabilities, as well as their electronic and vibrational behaviours, placing particular focus on $ep$ interactions and associated physical phenomena such as SC.

\section{Methods and Computational Details}
We investigated the phase space using the USPEX package for evolutionary crystal structure prediction~\cite{oganov2006crystal,lyakhov2013new} in the pressure range from 0 to \SI{250}{\giga\pascal}. For an overview, we started variable-composition runs in steps of \SI{50}{\giga\pascal}, providing already known phases as additional seed structures, and further refined our search to find phase boundaries within an \SI{5}{\giga\pascal} accuracy~\cite{a-USPEXdetails}.
We employed density functional theory (DFT) via the Quantum ESPRESSO (QE) package~\cite{giannozzi_quantum_2009} for all thermodynamically stable and metastable structures to relax the unit cells below a threshold of \SI{e-6}{Ry\per a_0} for all force components, and to calculate the electronic band structure and density of states (DOS) with an accuracy in the total energy between \SIrange{5}{10}{\milli\electronvolt/atom}~\cite{b-QEPWdetails}. We used scalar-relativistic optimized norm-conserving Vanderbilt pseudopotentials~\cite{hamann_optimized_2013,schlipf_optimization_2015}, a PBE+vdW functional~\cite{perdew_generalized_1996,thonhauser2007vdW} for calculations up to \SI{25}{\giga\pascal}~\cite{VDWdetails}, and a PBE-GGA functional for all calculations above \SI{25}{\giga\pascal}.
We calculated phonon dispersion relations, phonon DOSs as well as $ep$ coupling strengths $\lambda$ within the framework of density functional perturbation theory (DFPT), as implemented in QE~\cite{c-QEPHdetails}. The values for the critical temperature $T_c$ were estimated according to the Allen-Dynes McMillan formula~\cite{allen_transition_1975} with a typical value for the Morel-Anderson pseudopotential $\mu^*=0.1$, if not stated otherwise~\cite{morel_calculation_1962,mustar}. For \ce{Nb2S}, we additionally employed maximally localized Wannier functions and the fully anisotropic Migdal-Eliashberg theory as implemented in the EPW package to calculate $\lambda_{\mathbf{q},\nu}$, nesting functions $\zeta_{\mathbf{q}}$, and superconducting gaps $\Delta_{\mathbf{k}}(T)$ as a function of temperature $T$~\cite{ponce_epw:_2016,EPWdetails}.

\section{Phase Diagram}
\begin{figure}[t]
	\includegraphics[width=1.0\columnwidth]{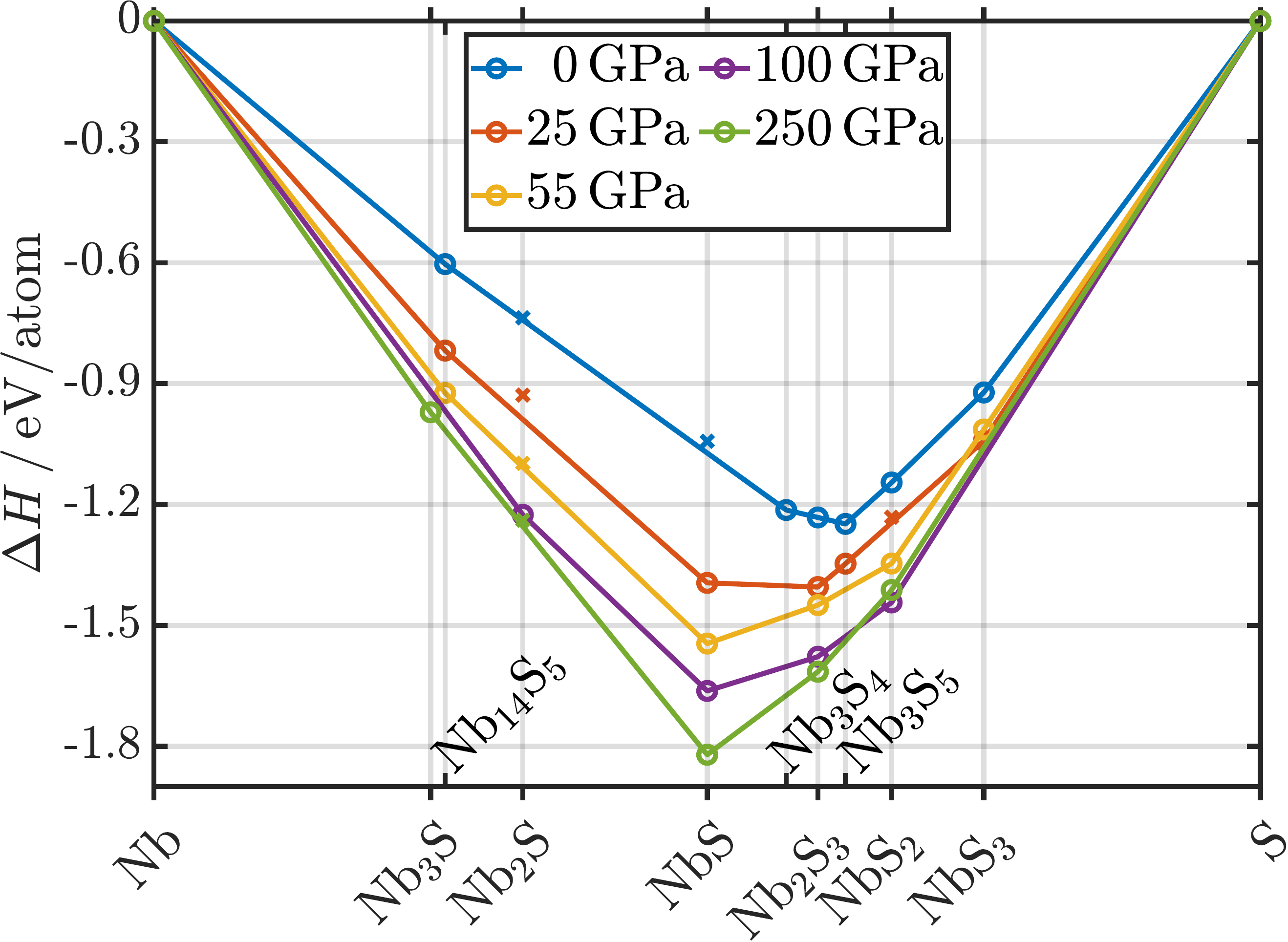}
	\caption{
		Convex hulls of enthalpy of formation $\Delta H$ for compounds $\mathrm{Nb}_x\mathrm{S}_y$ over sulfur fraction $y/(x+y)$ for different pressures. The circles mark stable structures, the crosses indicate a few selected metastable structures that are less than \SI{60}{\milli\electronvolt} above the convex hull.
	} 
	\label{fig:convex_hull}
\end{figure}

\begin{figure}[t]
	\includegraphics[width=1.0\columnwidth]{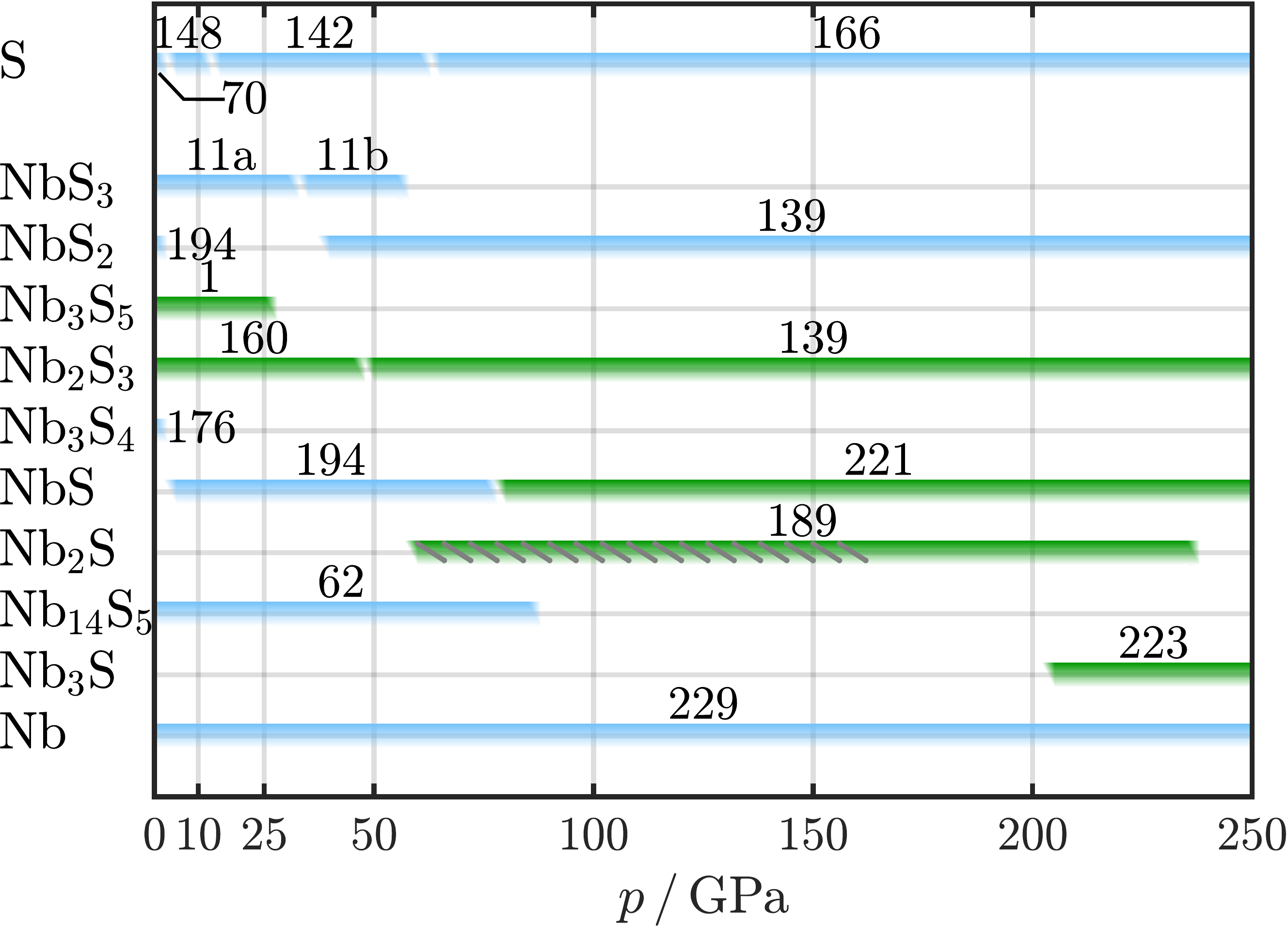}
	\caption{
		Phase diagram for the stable crystal structures in the Nb-S system. The numbers above the bars indicate the space group, the horizontal fading the transition pressures, and the colours previously reported (blue) and novel phases (green). The hatched region of \ce{Nb2S} denotes the dynamically unstable pressure range including anharmonic corrections.
	} 
	\label{fig:phase_diag}
\end{figure}

Following the evolutionary crystal searches we performed further relaxation and enthalpy calculations on the best candidates and determined convex hulls in the pressure range from ambient to \SI{250}{\giga\pascal} pressure, as shown in Fig.~\ref{fig:convex_hull} for five different pressures. A full phase diagram of stable structures and their space groups is presented in Fig.~\ref{fig:phase_diag}, where we indicate in blue previously reported phases and novel phases in green.  

In the studied pressure range, we reproduce the known phases of the Nb-S system in excellent agreement with literature~\cite{Jellinek1960NbS,Donohue1961SulfurS6,Smirnov1966Nb,ruysink1968Nb3S4,Chen1973Nb14S5,Meyer1976Sulfur,kikkawa1982NbS3,Zakharov1995bcoS,kenichi2006Nb,oganov2006crystal,liu2014NbS2,zhao2015Nb,kokail2016LiS,gavryushkin2017S,bloodgood2018NbS3,Debnath2018dissNbS3}: At ambient pressure the convex hull shows the stability of previously reported phases \ce{Nb14S5}, \ce{Nb3S4}, (2H-)\ce{NbS2}, and \ce{NbS3}, while \ce{Nb21S8} and \ce{NbS} observed in higher-temperature experiments~\mbox{\cite{franzen1968Nb21S8,SCHONBERG1954427,TaskinenNbSPhaseDiag,PredelNbSPhaseDiag}} appear as metastable, \SI{4}{\milli\electronvolt} and \SI{29}{\milli\electronvolt} above the convex hull tie-line, respectively. Increasing pressure, the formation of \ce{NbS} becomes favourable and it represents the convex hulls' minimum above \SI{30}{\giga\pascal} with a maximum enthalpy difference of about \SI{1.8}{\electronvolt/atom} at \SI{250}{\giga\pascal}.
 
Our structure searches also reveal several novel phases and materials of the Nb-S system, for which we provide all crystallographic details and schematics in the Supplemental Material (SM)~\cite{SMref}:
\begin {enumerate*} [label=\roman*)]
\item \ce{NbS} in the high-pressure \ce{CsCl}-type phase (space group 221),
\item \ce{Nb2S3} in different low- and high-pressure phases (space groups 160 and 139), appearing as ``doubled'' 3R-\ce{NbS2} and high\nobreakdash{-}pressure \ce{NbS2}, respectively (see Fig.~S2~\cite{SMref}), 
\item \ce{Nb3S5}, which appears as 3R-\ce{NbS2} intercalated with Nb atoms in octahedral coordination, resulting in the low-symmetry space group 1,
\item \ce{Nb3S}, a simple cubic material (space group 223) where the S atoms are in icosahedral coordination with the surrounding 12 Nb atoms, and the icosahedra being stacked in a bcc superlattice fashion,
\item three phases of \ce{Nb2S} (space groups 42, 11, and 189) that are described in detail in Sec.~\ref{sec:Nb2S}.
\end {enumerate*}

\begin{figure*}
	\includegraphics[width=1.0\textwidth]{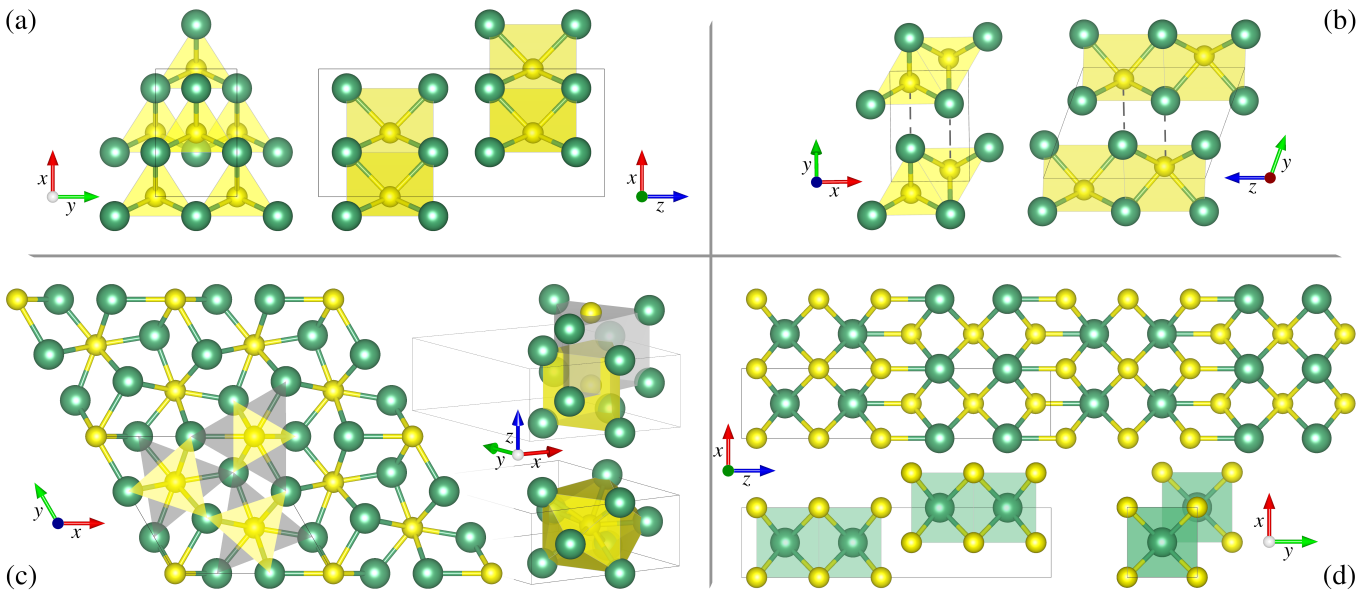}
	\caption{
		Characteristic crystal structures and geometries in the Nb-S system on the basis of \ce{Nb2S} and high-pressure \ce{Nb2S3}. Nb atoms are shown as large green spheres, S atoms as small yellow spheres, and the black solid lines indicate the unit cells.
		(a)~Metastable \ce{Nb2S} (space group 42) with S atoms in trigonal prismatic coordination with Nb shown from $z$-direction (left) and $y$-direction (right).
		(b)~Metastable \ce{Nb2S} (space group 11) shown from $z$-direction (left) and $x$-direction (right). This structure consists of a shifted stacking of low-pressure \ce{NbS} planes~\cite{SMref}, bringing certain Nb atoms in a distance to S atoms (dashed lines) that is only 10\% longer than the drawn bonds.
		(c)~High-pressure \ce{Nb2S} (space group 189) with S atoms in tricapped trigonal prismatic coordination with Nb. The coordination of a single S atom is highlighted in the lower 3D view, as a guide to the eye for the complex coordination geometry, a full (yellow) and an empty (grey) trigonal prism are shown in the upper right 3D view. A $2\times 2\times 1$ super cell is shown from $z$-direction on the left, where again the two prisms are indicated to highlight their hexagram orientation. They are shown for two S atoms inside the unit cell and one S atoms at the unit cell border, which are conceptually equal, but differ slightly in bond lengths.
		(d)~High-pressure \ce{Nb2S3} with a 2:2 stacking (see SM~\cite{SMref}). A $2\times 1\times 2$ super cell in $y$-direction is shown on top, the stacking and cubic coordination are emphasized in the bottom panels.
	}
	\label{fig:cs} 
\end{figure*}

Based on the types of appearing crystal structures, the phase diagram can be be separated roughly into low- and high-pressure regions between 50 and \SI{100}{\giga\pascal}. 
The low-pressure phases are predominantly built by layered \ce{NbS} units in the trigonal prismatic or octahedral geometry known from transition metal dichalcogenides~\cite{Manzeli2017TMDCryst,Leroux2018CDWNbS2CrystalStruct}, resulting in many layered (meta)stable structures with combined stacking patterns known from 1T-, 1H-, 2H- and 3R-\ce{NbS2} (see Figs.~S1 and S2~\cite{SMref}). On the other hand, the high-pressure sulfur-rich phases, i.e. $y/(x+y) \geq 0.5$ with $x$ being the number of Nb atoms per formula unit and $y$ the number of S atoms per formula unit, consist of different layered stackings of the \ce{CsCl}-like simple cubic NbS unit (space group 221) as illustrated in Fig.~\hyperref[fig:cs]{3d}.

For the high-pressure stacking with an even ratio \mbox{$n:n$}, i.e. $n$ simple cubic \ce{NbS} units on top of each other and $n$ units shifted by $(a/2,b/2,0)$ shown in Fig.~\hyperref[fig:cs]{3d} and Fig.~S3~\cite{SMref}, we find the stoichiometric formula $\mathrm{Nb}_n\mathrm{S}_{n+1}$ and the tetragonal space group 139, while an odd ratio \mbox{$n:(n+1)$} results in $\mathrm{Nb}_{2n+1}\mathrm{S}_{2n+3}$ and space group 123. Stable structures are predominantly formed with an even ratio and $n=1,2$, namely \ce{NbS2} and \ce{Nb2S3}, respectively, and metastable structures can be found with even and odd ratios up to $n=5$ (Fig.~S3~\cite{SMref}).

As widely reported in literature, the formation of layers is without doubt a common feature of transition metal chalcogenides and we therefore want to emphasize on some highly three-dimensional structures here:
\begin {enumerate*} [label=\roman*)]
\item \ce{Nb3S4}~\cite{ruysink1968Nb3S4,bullet1980NbS3Nb3S4}, in a hexagonal crystal structure, where all Nb atoms are in octahedral coordination, forming a 3D mesh-like structure.
\item The so far unknown \ce{Nb3S}, being the only Nb-rich member of the phase diagram at pressures above \SI{235}{\giga\pascal}. We also find an isostructural material with inverted empirical formula, \ce{NbS3}, that is metastable at high pressures.
\item  The so far unknown high-pressure structure of \ce{Nb2S}, where nine Nb atoms forming a 14-faced structure around the S atoms are three-dimensionally packed in a hexagonal lattice.
\end {enumerate*}
The crystal structure of \ce{Nb2S}, which we will discuss in detail in Sec.~\ref{sec:Nb2S}, as well as examples of the characteristic crystal geometries in the Nb-S system are shown in Fig.~\ref{fig:cs}. 

For all materials not extensively discussed in literature we carried out electronic, vibrational, and $ep$ calculations, revealing that all phases except pure sulfur at low pressures are metallic and superconducting with $T_c < \SI{15}{\kelvin}$. To not deter from the main focus of this paper, we provide an extensive discussion of all materials and their properties in the SM~\cite{SMref} and in the following analyse in detail the novel \ce{Nb2S} compound, for which we find imaginary phonon modes within DFPT at certain parts of the Brillouin zone (BZ) that are indicative of lattice instabilities. In the next section we will show that this putative instability is due to the disregard of anharmonic effects and that considering the full anharmonic potential, \ce{Nb2S} is in fact stable over a wider pressure range exhibiting superconductivity and strong $ep$ coupling.

\section{\texorpdfstring{$\textbf{\ce{Nb2S}}$}{Nb2S} - Anharmonicity and Superconductivity}
\label{sec:Nb2S}
At low pressures, we find two metastable phases of \ce{Nb2S} (see convex hulls in Fig.~\ref{fig:convex_hull}), namely an orthorhombic structure (space group 42) at ambient pressure that consists of vdW-coupled layers in trigonal prismatic geometry around sulfur atoms~\cite{note_inverted}, shown in Fig.~\hyperref[fig:cs]{3a}, and above \SI{20}{\giga\pascal} a monoclinic structure (space group 11) with a different trigonal prismatic arrangement exhibiting a denser clustering of \ce{Nb} atoms around \ce{S} atoms (indicated by dashed lines in Fig.~\hyperref[fig:cs]{3b}). The former lies  \SI{6}{\milli\electronvolt} above the convex hull tie-line, and the latter \SI{60}{\milli\electronvolt}. Above \SI{40}{\giga\pascal}, a hexagonal phase (space group 189) with nine atoms per unit cell becomes more favourable, but appears on the convex hull only above \SI{60}{\giga\pascal} due to the presence of \ce{Nb14S5}. Here, the main repetition unit is a non-uniform 14-faced triaugmented triangular prism around S atoms, formed by two shifted and rotated triangular prisms in hexagram geometry as illustrated in Fig.~\hyperref[fig:cs]{3c}. As indicated by the shaded areas in the supercell, the side lengths of the two auxiliary trigonal prisms (yellow and grey) are different, as well as the side lengths of the conceptually equal prisms around S atoms inside the unit cell and around S atoms at the unit cell border. Above \SI{235}{\giga\pascal} the simple cubic \ce{Nb3S} structure becomes thermodynamically more favourable and \ce{Nb2S} leaves the convex hull. 

\begin{figure}[t]
	\includegraphics[width=1.0\columnwidth]{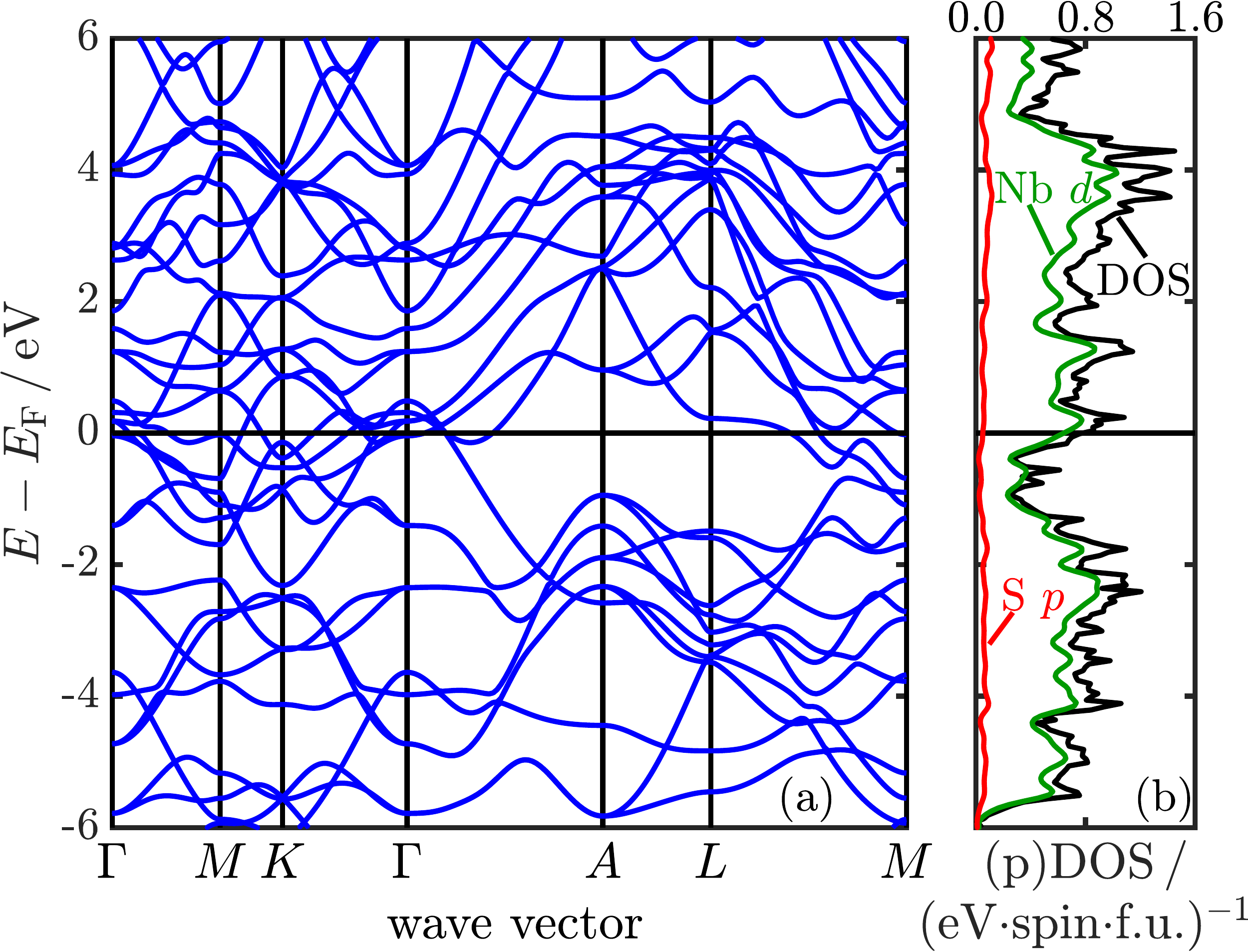}
	\caption{
		Electronic properties of \ce{Nb2S} at \SI{225}{\giga\pascal}. (a) The band structure around the Fermi level along a high-symmetry BZ path and (b) the (partial) DOS and orbital contributions. The green curve shows the sum over all Nb $d$ orbitals, the red curve the sum over all S $p$ orbitals.
	} 
	\label{fig:Nb2S_el_225}
\end{figure}

In Fig.~\ref{fig:Nb2S_el_225}, we show the electronic band structure around the Fermi level of the high-pressure \ce{Nb2S} phase along a high-symmetry path of the hexagonal BZ. We find highly anisotropic dispersions in all directions, supporting our structural observations that this is indeed a highly three-dimensional material. The DOS together with the partial projections onto the Nb $d$ and S $p$ orbitals are reported in the same figure. We find the Fermi level on the lower shoulder of a peak in the DOS, a property that remains robust over the stable pressure range, as demonstrated in Fig.~S4~\cite{SMref}. In the energy range around and especially at the Fermi energy, the main contributions to the total DOS originate from the Nb $d$ orbitals, a prominent feature already observed in group-5 TMDs~\cite{zhou2017tetTMC} and also present in most other phases in the Nb-S system~\cite{SMref}.

\begin{figure}[t]
	\includegraphics[width=1.0\columnwidth]{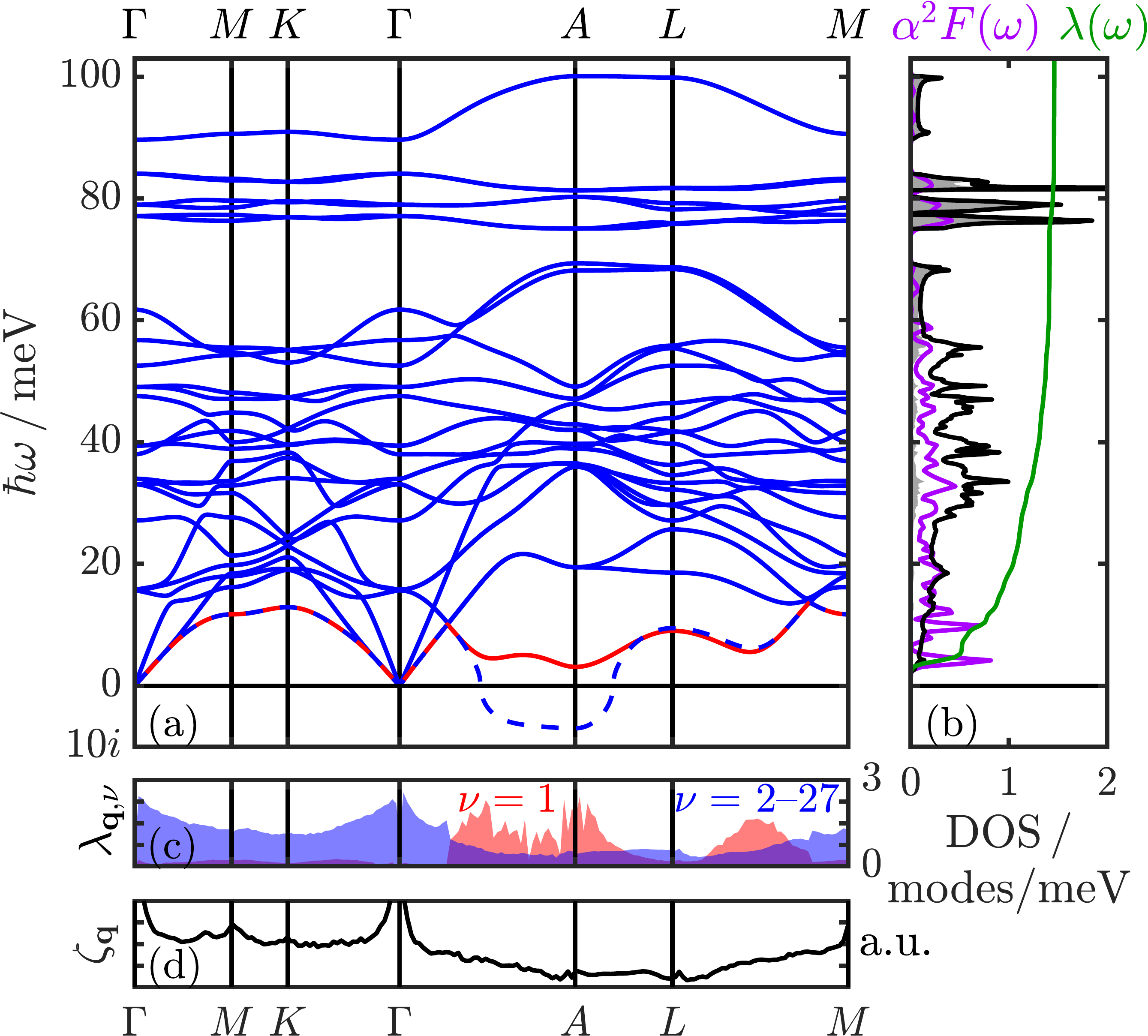}
	\caption{
		Vibrational and $ep$ properties of \ce{Nb2S} at \SI{225}{\giga\pascal}. (a) Phonon dispersion, where the dashed line represents the imaginary mode as given by the harmonic approximation and the red solid line indicates the full anharmonic result. (b) Phonon DOS (black line) with S (Nb) contributions as grey (white) area, Eliashberg function $\alpha^2F(\omega)$ (purple), and cumulative $ep$ coupling strength $\lambda(\omega)$ (green). (c) mode-resolved $ep$ interaction $\lambda_{\mathbf{q},\nu}$ along the high-symmetry path for the lowest (red shaded area, $\nu=1$) and all other modes (blue shaded area). (d) nesting function $\zeta_{\mathbf{q}}$ along the same path. 
	} 
	\label{fig:Nb2S_ph_225}
\end{figure}

Vibrational properties of high-pressure \ce{Nb2S} are shown in Fig.~\ref{fig:Nb2S_ph_225}. The main contribution to the phonon DOS up to the peak below \SI{60}{\milli\electronvolt} stems from the heavier Nb atoms, and above that from the lighter S atoms, as indicated by the white and grey areas underneath the DOS curve in Fig.~\hyperref[fig:Nb2S_ph_225]{5b}. The S-dominated DOS range is separated by two phonon band gaps around 70 and \SI{85}{\milli\electronvolt}. In the harmonic phonon dispersion (blue lines in Fig.~\hyperref[fig:Nb2S_ph_225]{5a}) we find a single soft phonon mode exhibiting imaginary phonon frequencies in a BZ region around $\mathbf{q}_1 = (0,0,1/2) = A$, clearly visible along the path $\Gamma-A-L$, and suggesting an apparent lattice instability. As demonstrated in Ref.~\cite{heil2017NbS2} for the case of \ce{NbS2} though, imaginary (harmonic) phonon frequencies are not a conclusive indication for lattice instabilities, but rather require a deeper quantum mechanical treatment. For that purpose, we calculated explicitly the corresponding adiabatic potential energy surface (APES) for those $\mathbf{q}$-points that have imaginary phonon modes. This was done via a frozen phonon approach, for which we constructed appropriate supercells in order to fold a specific $\mathbf{q}$-point back to $\Gamma$, where the DFPT solution provides a real phonon eigenvector and therefore directly the atomic displacements. 
In our case, where we find imaginary modes at $\mathbf{q}_1 = A$ and $\mathbf{q}_2 = 2/3\,A$, this leads to $1\times 1\times 2$ and $1\times 1\times 3$ supercells, respectively.
We performed total energy calculations as a function of the phonon eigenvector amplitude, allowing us to construct the full anharmonic APES, for which we then solved the one-dimensional Schr\"odinger equation~\cite{SM_2D}. 
To calculate phonon dispersion relations and electron-phonon interactions in the presence of anharmonicity, we retain the (harmonic) DFPT eigenmodes and use the anharmonically corrected frequencies, approximated via the difference between the first two eigenenergies of the fully anharmonic APES, to compute the interatomic force constants and dynamical matrices~\cite{SM_PRL}.
We also find that taking into account the full anharmonic potential leads to ground state probability densities centered at the high-symmetry structures, i.e. non-displaced atoms, and non-vanishing anharmonic phonon frequencies, and therefore to a stabilization of the \ce{Nb2S} structure. The anharmonically corrected phonon dispersion is presented in Fig.~\hyperref[fig:Nb2S_ph_225]{5a} using the solid red line. 

In Fig.~\hyperref[fig:Nb2S_ph_225]{5b} we report the Eliashberg spectral function $\alpha^2F(\omega)$ and cumulative $ep$ coupling strength $\lambda(\omega)$, including  anharmonic corrections. The Eliashberg function essentially follows the phonon DOS for energies above \SI{15}{\milli\electronvolt}, but exhibits a few additional peaks in the low-energy range corresponding to the flat dispersion of the anharmonic mode along the path $L-M$. In accordance with the sharp increases in the cumulative $\lambda$ due to the peaks in $\alpha^2F(\omega)$, we find that the low-frequency, anharmonic mode contributes around 50\% to the total $ep$ coupling of about $\lambda=1.5$. In order to elucidate the origin of this strong $ep$ interaction, we calculated the mode- and wave vector resolved $ep$ coupling strength $\lambda_{\mathbf{q},\nu}$. As expected, we find a large portion of the total $ep$ coupling stemming from the anharmonic mode around $A$ and close to $L-M$~(Fig.~\hyperref[fig:Nb2S_ph_225]{5c}). 
Conversely, we observe no strong response at these regions of the BZ for the nesting function $\zeta_\mathbf{q}$~(Fig.~\hyperref[fig:Nb2S_ph_225]{5d}), indicating that the observed anharmonicity and softening of phonon modes is purely due to a strong, spatially localized $ep$ coupling. 

\begin{figure}[t]
	\includegraphics[width=1.0\columnwidth]{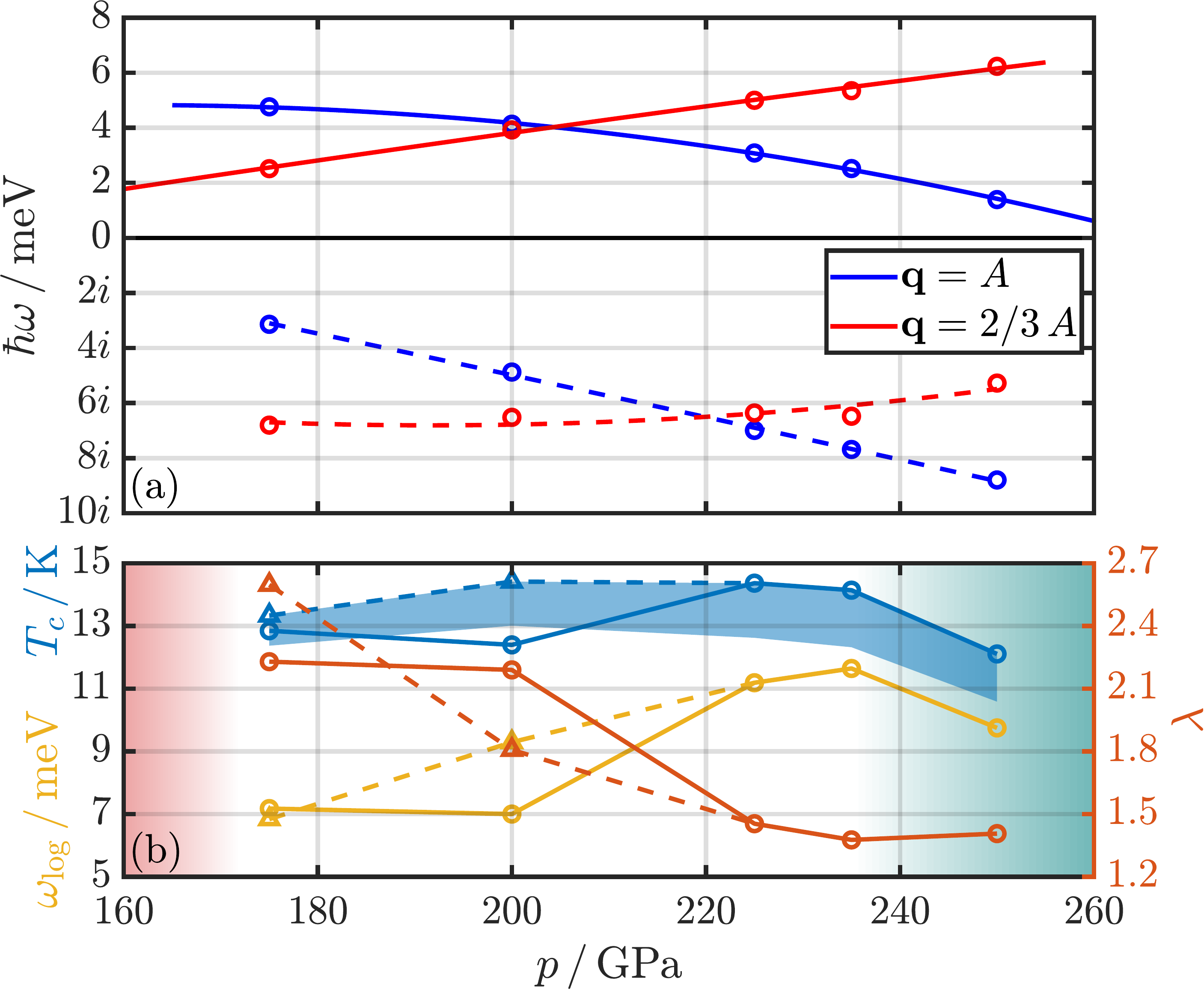}
	\caption{
	(a)	Anharmonic phonon frequencies (top panel) and harmonic approximations (lower panel) for the soft mode at $\mathbf{q}_1=A$ and $\mathbf{q}_2=2/3\,A$ in the pressure range from \SI{175} to \SI{250}{\giga\pascal}. The solid and dashed lines are quadratic fits as guides to the eye. 
	(b) Electron-phonon related quantities $\lambda$, $\omega_\text{log}$, and $T_c$ as a function of pressure. The shaded area for $T_c$ corresponds to different choices of $\mu^*$, with the largest values at a given pressure corresponding to $\mu^*=0.1$ and the smallest to $\mu^*=0.14$. The triangles at 175 and \SI{200}{\giga\pascal} correspond to calculations where the soft mode frequency at $\mathbf{q}_3$ and $\mathbf{q}_4$ has been shifted ``by hand'' (see text). The dashed and solid lines are again guides to the eye. The shaded area below \SI{175}{\giga\pascal} indicates the dynamic instability and the shaded area above \SI{235}{\giga\pascal} indicates the thermodynamic instability.
	} 
	\label{fig:anharm_over_p}
\end{figure}

In Fig.~\hyperref[fig:anharm_over_p]{6a}, we trace the anharmonic modes for the two irreducible wave vectors $\mathbf{q}_1$ and $\mathbf{q}_2$ over pressure and find an opposed behaviour, i.e., the frequency of the $\mathbf{q}_2$ mode increases with pressure, while it decreases for $\mathbf{q}_1$. Extrapolating the energy of the $\mathbf{q}_1$ mode, we find a hypothetical dynamic stability limit around \SI{265}{\giga\pascal}. (We want to note at this point that \ce{Nb3S} already becomes thermodynamically more favourable for pressure above \SI{240}{\giga\pascal}.) An extrapolation of the results for $\mathbf{q}_2$ shows that the energy of this mode will vanish for pressures below \SI{130}{\giga\pascal}.

Decreasing the pressure, we also find that the BZ regions of imaginary harmonic modes increase, extending ever closer to and eventually including $\Gamma$ and additionally appearing on the path $\Gamma-L$, as shown in Fig.~S6~\cite{SMref}. At \SI{150}{\giga\pascal} and below, the ground state probability density of the soft mode at $\mathbf{q}_2$ is no longer centered around the high-symmetry structure, but has its maxima in the minima of the double-well APES, thus indicating the transition to a different structure with lower symmetry.

Based on our phonon dispersions for various pressures, we do expect that for pressures between \SI{175}{\giga\pascal} and \SI{200}{\giga\pascal}, \ce{Nb2S} would still be dynamically stable when anharmonicity would be included for all wave vectors. As the corresponding supercells would be very large and the calculations therefore computationally prohibitively expensive, however, we did not correct the additionally appearing imaginary modes below \SI{200}{\giga\pascal} by a full calculation of the APES. Instead, we chose to approximately investigate the sensitivity of the results with respect to anharmonic effects by shifting the soft mode energies of the additional wave vectors ``by hand'': The choice of our $\mathbf{q}$-grid results in two irreducible $\mathbf{q}$-points in these regions, namely $\mathbf{q}_3=1/3\,A$ and $\mathbf{q}_4=(0,1/3,1/3) = 2/3\,L$, for which we set the modes' energies such as to achieve fully positive dispersions in the whole BZ at \SI{175}{\giga\pascal} and \SI{200}{\giga\pascal}, and no soft mode energy below \SI{1}{\milli\electronvolt} in order to avoid artificially high $ep$ couplings.

This allows us to evaluate the superconducting properties over the full stable pressure range of \ce{Nb2S}.
In Fig.~\hyperref[fig:anharm_over_p]{6b}, we show the total $ep$ coupling strength $\lambda$, $\omega_{\mathrm{log}}$, and $T_c$ as functions of pressure. The circles and solid lines correspond to calculations for which anharmonic effects are only considered for wave vectors $\mathbf{q}_1$ and $\mathbf{q}_2$ (all $ep$ contributions from imaginary phonon modes at $\mathbf{q}_3$ and $\mathbf{q}_4$ are set to zero), while shifting the frequencies of $\mathbf{q}_3$ and $\mathbf{q}_4$ ``by hand'' as described in the previous paragraph leads to the triangles and dashed curves. The latter yield smoother and dome-shaped curves for $T_c$, supporting our initial hypothesis that the APES associated with the phonon modes at $1/3\,A$ and $2/3\,L$  have indeed a strong anharmonic part.

We find $T_c$ values ranging between \SI{12} and \SI{14}{\kelvin} (see Tab.~S1 in the SM for a comparison with the other materials in the Nb-S system~\cite{SMref}). $\lambda$ increases dramatically when decreasing pressure due to increasing anharmonic effects, reaching values comparable to the current $T_c$ record holders in the high-pressure hydride class~\cite{duan2014pressure,heil_influence_2015,flores_SC_2016,Liu_potential_2017,Drozdov2019LaH10,Heil_superconductivity_2019,errea2020quantum}. The characteristic phonon frequency $\omega_{\mathrm{log}}$, however, is dominated by low-frequency modes of heavy Nb atoms, thus counteracting the high $ep$ coupling, resulting in only a moderate critical temperature $T_c$.

We further performed calculations in the framework of the fully anisotropic Migdal-Eliashberg theory, in order to obtain the superconducting gap as function of temperature, and find an isotropic and single-gap distribution of the superconducting gap, as detailed in Fig.~S5~\cite{SMref}. 

\section{Conclusions}
In this work we studied the binary phase diagram of the Nb-S system and investigated $ep$ coupling and superconducting properties of its phases up to \SI{250}{\giga\pascal}, using fully \textit{ab initio} methods. Our results are in excellent agreement with literature for previously reported phases and we find several new materials, revealing an intricate phase diagram at all pressures. Apart from crystal structures consisting mainly of layers with distinct high- and low-pressure building blocks, we also find phases with particularly strong three-dimensional character, and except for the low-pressure phases of pure sulfur, all investigated structures are superconducting metals with $T_c$'s below \SI{15}{\kelvin}. 

We focused in particular on the highly three-dimensional, high-pressure phase of \ce{Nb2S} that exhibits an apparent lattice instability. By taking into account the full anharmonic potential energy surface, however, we show that this instability is suppressed, demonstrating that imaginary harmonic phonon frequencies are not a conclusive indicator for lattice instabilities, but require a more sophisticated quantum-mechanical treatment. By ruling out Fermi surface nesting, the origin of the strong anharmonicity is found to be in most parts due to an $ep$ interaction particularly strong for a single phonon mode and certain wave vector regions of the Brillouin zone.

Employing the fully anisotropic Migdal-Eliashberg theory, we find that the superconducting gap function is fairly isotropic with only a single gap. We further investigate the influence of the anharmonic corrections on $ep$ coupling and superconductivity, as well as their dependence on pressure, and find a dome-shaped behaviour for $T_c$ with a maximum of \SI{14.4}{\kelvin} at \SI{225}{\giga\pascal}.

Our research reveals a much more complex phase diagram for the Nb-S system than indicated by available literature, featuring not only the characteristic, layered vdW structures, but also highly three-dimensional phases at low and high pressures, some of which exhibit interesting anharmonic phonon behaviour and strong $ep$ interaction. Shedding new light on the Nb-S system, our findings encourage to revisit familiar systems with the promise of discovering novel and fascinating phases.

\section*{acknowledgments}
This work was supported by the Austrian Science Fund (FWF) Project No. P 32144-N36, the dCluster of the Graz University of Technology, and the VSC4 of the Vienna University of Technology.

%\bibliographystyle{apsrev4-2}
%\bibliography{paper}{}

\begin{thebibliography}{100}
\bibitem [{a-U()}]{a-USPEXdetails}%
	\BibitemOpen
	\href@noop {} {}\bibinfo {note} {The variable-composition runs were executed
		for 25 generations with a population size of 50 individuals each. The new
		generations were created based on 60\% of the best individuals in the old
		generation. The relaxations were done in a 5-step procedure with increasing
		$\mathbf{k}$ resolution of $|\Delta\mathbf{k}| = 0.15 - \SI{0.07}{\cdot
			2\pi\per\angstrom}$. We performed several runs with minimum 4 to maximum 24
		atoms per unit cell (uc). Materials with more than 24 atoms/uc were not
		considered, apart from previously reported \ce{Nb14S5} (76 atoms/uc) and
		\ce{Nb21S8} (29 atoms/uc) that were added as seeds by hand.}\BibitemShut
	{Stop}%
\bibitem [{b-Q()}]{b-QEPWdetails}%
	\BibitemOpen
	\href@noop {} {}\bibinfo {note} {We performed extensive convergence tests on
		$\mathbf{k}$-point spacing, kinetic energy cutoff and smearing value for the
		Brillouin zone integration. Converged values were obtained with
		$\left|\Delta\mathbf{k}\right|=\SI{0.01}{\cdot 2\pi\per\angstrom}$, a
		smearing of \SI{0.01}{Ry}, and an energy cutoff of \SI{125}{Ry}. We used a
		convergence threshold of \SI{e-10}{Ry} for the total energy in all electronic
		self-consistence calculations. The thresholds for the relaxation calculations
		were \SI{e-7}{Ry} in total energy and \SI{e-6}{Ry\per a_0} for all force
		components.}\BibitemShut {Stop}%
\bibitem [{VDW()}]{VDWdetails}%
	\BibitemOpen
	\href@noop {} {}\bibinfo {note} {The differences in the convex hulls
		calculated with and without vdW corrections range from
		\SIrange{50}{100}{\milli\electronvolt} at ambient pressure, but are below
		\SI{10}{\milli\electronvolt} at \SI{25}{\giga\pascal}}\BibitemShut {NoStop}%
\bibitem [{c-Q()}]{c-QEPHdetails}%
	\BibitemOpen
	\href@noop {} {}\bibinfo {note} {The phonon convergence tests were based on
		reference calculations with a kinetic energy cutoff of \SI{200}{Ry},
		$\left|\Delta\mathbf{k}\right|=\SI{0.003}{\cdot 2\pi\per\angstrom}$ and a
		phonon self-convergency threshold of \SI{e-18}{}. The results were considered
		converged if the first optical mode's frequency at $\mathbf{q}=\Gamma$ was
		within \SI{1}{\milli\electronvolt} or smaller compared to the reference
		calculation. We obtained converged results for a kinetic energy cutoff for
		the plane waves of \SI{60}{Ry},
		$\left|\Delta\mathbf{k}\right|=\SI{0.013}{\cdot 2\pi\per\angstrom}$ and
		phonon self-convergency threshold of \SI{e-14}{}. The phonon dispersions were
		calculated with a $\left|\Delta\mathbf{q}\right|=\SI{0.05}{\cdot
			2\pi\per\angstrom}$. The electron-phonon properties were calculated with a
		$\left|\Delta\mathbf{k}\right|=\SI{0.006}{\cdot 2\pi\per\angstrom}$ using
		broadening values between \SI{5e-3} and \SI{5e-2}{Ry}, and a phonon smearing
		for $\alpha^2F$ of \SI{0.15}{\tera\hertz}.}\BibitemShut {Stop}%
\bibitem [{mus()}]{mustar}%
	\BibitemOpen
	\href@noop {} {}\bibinfo {note} {In order to estimate the
				effect of $\mu^*$ on $T_c$, we have calculated $T_c$ for Nb$_2$S at various
				pressures using the Allen-Dynes McMillan equation for several values $\mu^*$ up to 0.14, resulting in a maximal total variation of around 2~K (see Fig.~\ref{fig:anharm_over_p}). In addition, we have also repeated our Migdal-Eliashberg calculations for Nb$_2$S at 225~GPa pressure with $\mu^*=0.14$, leading to a
				change in $T_c$ also by $\sim 2$~K (see Fig.~S5 of the SM~\cite{SMref}).}\BibitemShut
	{Stop}%
\bibitem [{EPW()}]{EPWdetails}%
	\BibitemOpen
	\href@noop {} {}\bibinfo {note} {We used projections onto Nb $p,d$ and S $p$
		for the Wannierization, and a coarse $\mathbf{k} (\mathbf{q})$-grid of $6
		\times 6 \times 12$ ($3 \times 3 \times 6$). For the nesting function and
		$\lambda_\mathbf{q}$, we interpolated over a fine $\mathbf{k}$-grid of $36
		\times 36 \times 72$, and for the Migdal-Eliashberg calculations we employed
		fine $\mathbf{k}/\mathbf{q}$-grids of $12 \times 12 \times 24$.}\BibitemShut
	{Stop}%
\bibitem [{SM_({\natexlab{a}})}]{SM_2D}%
	\BibitemOpen
	\href@noop {} {}\bibinfo {note} {The
				approach adopted here includes by construction phonon-phonon interactions
				within the same harmonic eigenmode, but it neglects those between different
				eigenmodes. This coupling could be included by considering multi-dimensional
				APES, for which the atomic displacements are expanded in terms of multiple
				phonon eigenmodes. To do this for all phonon modes is computationally
				prohibitive, which is why we tested the coupling between the two lowest
				phonon modes at $A$ and $2/3\,A$ by calculating the two-dimensional APES, as
				shown in Fig.~S27 and Fig.~S28 of the SM~\cite{SMref}. This analysis shows
				that the coupling between the first and the second phonon modes at $A$ and
				$2/3\,A$ is small, and that it can, in good approximation, be neglected for the purposes of this work.}\BibitemShut {Stop}%
\end{thebibliography}

\end{document}

\title{Supplemental material\texorpdfstring{\\}{-} Superconductivity and strong anharmonicity in novel Nb-S phases}
	
\author{Roman Lucrezi} 
\affiliation{\Graz}
\author{Christoph Heil} \email{christoph.heil@tugraz.at}
\affiliation{\Graz}

\date{January 11, 2021}

\maketitle

\section{Thermodynamic stability}
We determined the thermodynamically stable $\ce{Nb}_x\ce{S}_y$ phases $(\pi_{x,y})$ at a fixed external pressure $p$ via their enthalpy of formation
\begin{equation}
 \Delta H(\pi_{x,y}) = H(\pi_{x,y}) - \frac{xH(\pi_{1,0})+yH(\pi_{0,1})}{x+y} 
\end{equation}
with respect to the boundary phases of niobium $\pi_{1,0}$ and sulfur $\pi_{0,1}$, where 
\begin{equation}
 H(\pi_{x,y}) = \frac{E_{\mathrm{tot}}(\pi_{x,y})}{N} + p\frac{V(\pi_{x,y})}{N}
\end{equation}
is the enthalpy per atom, $E_{\mathrm{tot}}$ the total energy, $N$ the number of atoms in the unit cell, and $V$ the unit-cell volume. The sets of phases that are additionally stable against decomposition into neighbouring phases form so-called \textit{convex hulls} for a fixed pressure, as shown in Fig.~1 of the main text.

\section{Comments on previously reported phases}
\begin{itemize}
 \item Sulfur is an element with many known allotropes. 
 The most stable phases in our calculations are the well known $\alpha$-S (\ce{S8} rings in space group 70) at ambient pressure, \ce{S6} rings from \SIrange{5}{20}{\giga\pascal} (space group 148), spiral tetragonal chains in space group 142 from \SIrange{20}{70}{\giga\pascal}, and the $\beta$-\ce{Po} phase (space group 166) above \SI{70}{\giga\pascal} \cite{gavryushkin2017S,oganov2006crystal,kokail2016LiS}. We also found and included structures with trigonal spiral chains in one and three orientations~\cite{oganov2006crystal}, as well as other orientations of \ce{S8}, but they did not appear as thermodynamically stable phases in our calculations. The same applies to an incommensurate base-centered orthorhombic (bco) S, whose stability has not been confirmed by theoretical calculations~\cite{Degtyareva2005bcoS,Akahama1993bcoS,kokail2016LiS}.
 We want to note at this point that we found some structures in the transition range from space group 142 to 166 (\SIrange{60}{75}{\giga\pascal}) formally lower in enthalpy, but with an enthalpy difference smaller than our accuracy of \SI{5}{\milli\electronvolt}. These structures show some similarity to the puckered layers proposed for
 the bco S modelling \cite{Zakharov1995bcoS}, yet with lower symmetry. 
 \item We find all known low-pressure phases of \ce{NbS2} (1T-, 1H-, 2H- and 3R) and especially confirm the extensively studied 2H structure as the stable phase at ambient pressure \cite{heil2017NbS2,tissen2013NbS2}.
 We also confirm the high-pressure phase (space group 139), although we find a slightly higher transition pressure as predicted in Ref.~\cite{liu2014NbS2} (between \SI{30}{\giga\pascal} and \SI{35}{\giga\pascal} instead of \SI{26}{\giga\pascal}) that can be explained by taking into account vdW as shown in Ref.~\cite{Wang2020NbS2}, where a transition pressure of \SI{38}{\giga\pascal} is predicted. We further find imaginary phonon frequencies at $\mathbf{q}=X$ below \SIrange{100}{150}{\giga\pascal}, which were not predicted in Ref.~\cite{liu2014NbS2}.
 \item \ce{NbS3} is known in several phases usually (historically) labelled I to V and ``HP'' (for high pressure, referring to a <\SI{10}{\giga\pascal} range). I, IV and V are \ce{ZrSe3}-like structures (space group 11 with 8 atoms in the unit cell) with different distortions, where V is the most symmetric, I is slightly distorted and the distortion in IV reaches over two unit cells. While \ce{NbS3}-I is commonly considered to be the predominant structure formed in experiments, our calculations yield the larger \ce{NbSe3}-like structure ``\ce{NbS3}-HP'' (24 atoms in the unit cell) to be thermodynamically stable up to \SI{30}{\giga\pascal} (labelled ``\ce{NbS3} a'' in the phase diagram in Fig.~2 of the main text). From \SIrange{30}{55}{\giga\pascal}, \ce{NbS3}-V is more favourable (labelled ``\ce{NbS3} b''), for which we confirm imaginary frequencies in different BZ regions and strong $ep$ couplings $\lambda>2$ \cite{rijnsdorp1978NbS3,bloodgood2018NbS3,srivastava1992trichalco,Debnath2018dissNbS3}.
\end{itemize}

\section{Low-pressure geometries}
\begin{figure}[ht]
	\includegraphics[width=0.6\textwidth]{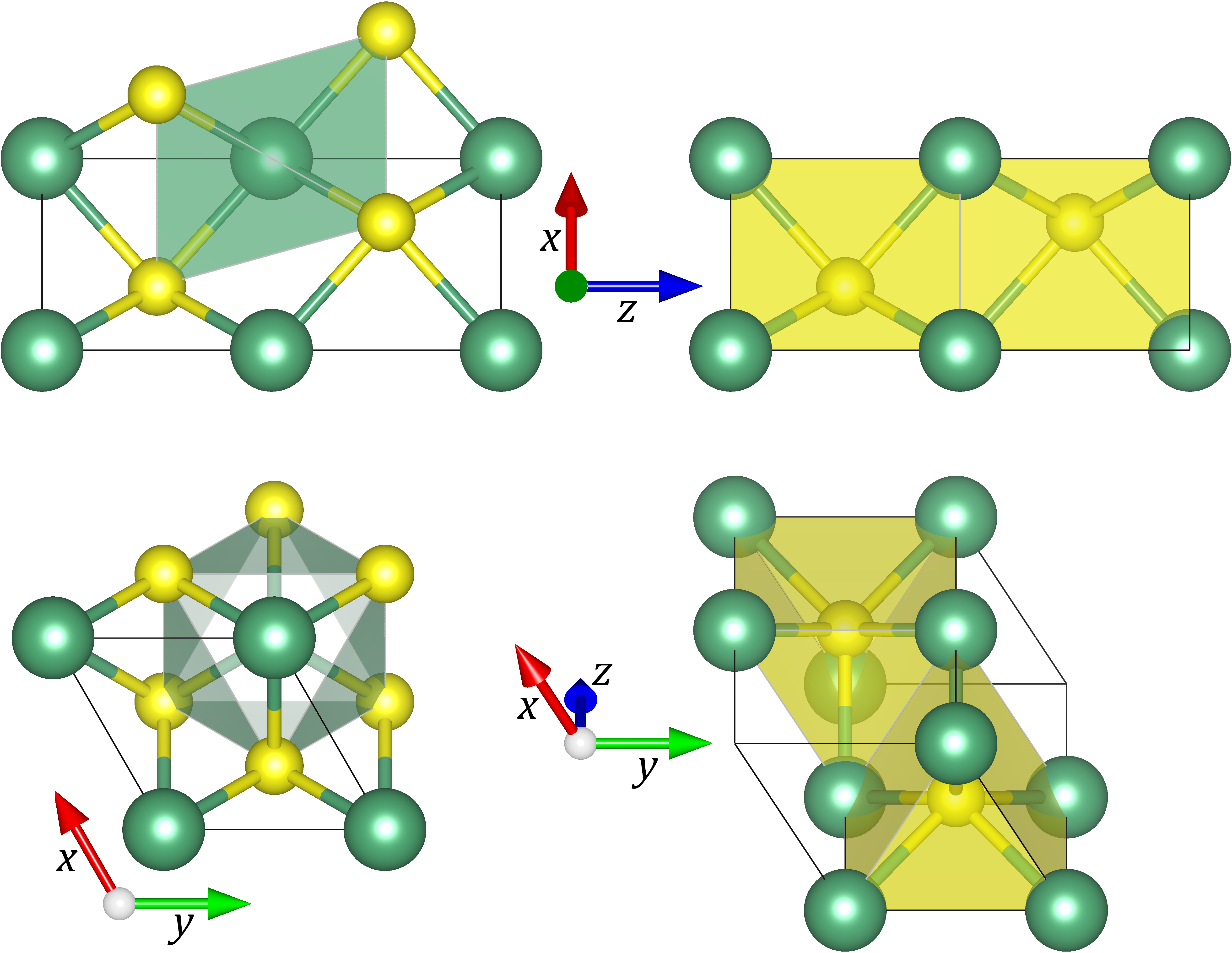}
	\caption{
		Low pressure \ce{NbS} (space group 194): Nb in octahedral coordination on the left, S in trigonal prismatic coordination on the right.
	} 
	\label{fig:NbS_coord}
\end{figure}

The low-pressure structures in the Nb-S system are dominated by trigonal prismatic or octahedral coordination geometries possible for both element types. The coordination number for both geometries is six, stemming from two atomic triangles that are either aligned (prism) or rotated by \SI{60}{\degree} (octahedron). Figure~\ref{fig:NbS_coord} shows the coordination geometries of Nb and S atoms in the low-pressure \ce{NbS} phase (space group 194).

\begin{figure}[ht]
	\includegraphics[width=\textwidth]{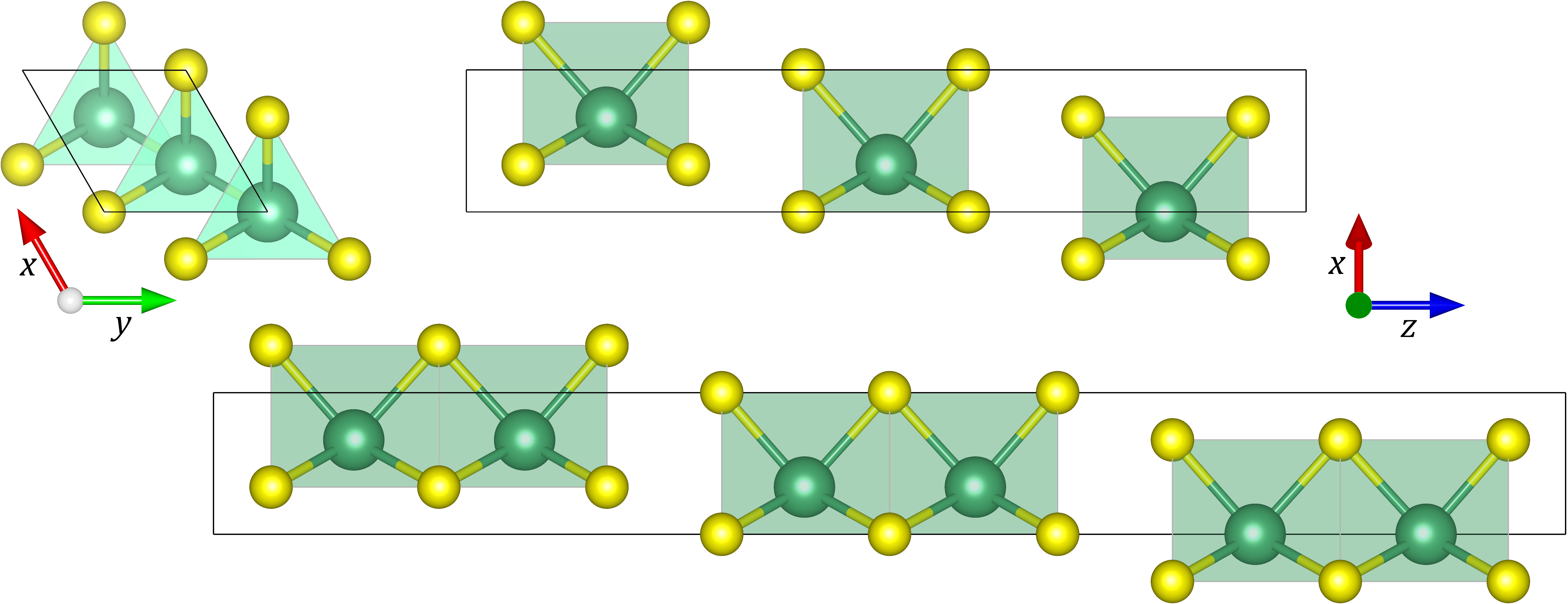}
	\caption{
		Low pressure \ce{Nb2S3} (lower) with trigonal prisms stacked in the pattern from metastable 3R-\ce{NbS2} (upper). Both structures belong to space group 160.
	} 
	\label{fig:Nb2S3_3RNbS2}
\end{figure}

In many metastable phases, we find the coordination geometries in combination with stacking patterns known from other \ce{NbS2} phases. For example, we report a stable phase for \ce{Nb2S3} formed in that fashion from 3R-\ce{NbS2} by doubling the individual vdW layers as shown in Fig.~\ref{fig:Nb2S3_3RNbS2}.

\section{High-pressure stacking}
\begin{figure}[ht]
	\includegraphics[width=\textwidth]{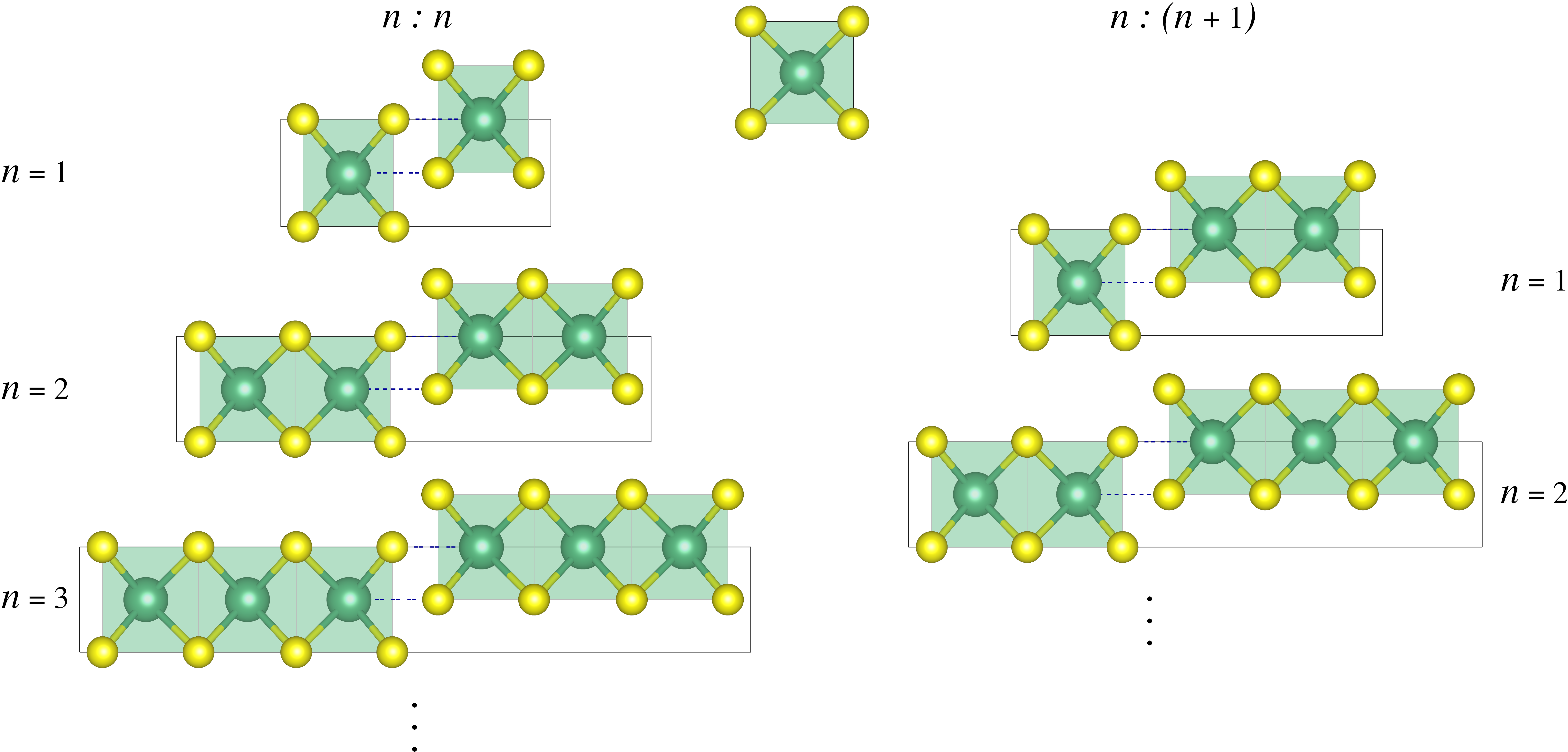}
	\caption{
		Stacking patterns of the simple cubic \ce{NbS} (center) with even $n:n$ ratios on the left and odd $n:(n+1)$ ratios on the right. Even structures include \ce{NbS2} ($n=1$), \ce{Nb2S3} ($n=2$), and \ce{Nb3S4} ($n=3$). Odd structures include \ce{Nb3S5} ($n=1$) and \ce{Nb5S7} ($n=2$).
	} 
	\label{fig:hp_NbS_stacking}
\end{figure}

The high-pressure structures in the Nb-S system are dominated by diagonal stackings of simple cubic high-pressure \ce{NbS} units, forming structures of space group 139 and 123 for even and odd stacking ratios, respectively. Figure \ref{fig:hp_NbS_stacking} illustrates the stacking patterns.

Even ratios $n:n$ result in structures with stoichiometric formula $\mathrm{Nb}_n\mathrm{S}_{n+1}$, and odd ratios $n:(n+1)$ in structures $\mathrm{Nb}_{2n+1}\mathrm{S}_{2n+3}$. We found many metastable structures up to $n=5$, but only the even stacking for $n=1$ and $n=2$ resulted in the thermodynamically stable phases of \ce{NbS2} and \ce{Nb2S3}. For the sake of curiosity, we constructed structures with different even stackings, i.e. $n:(n+2)$, resulting in the same $n:n$ formula, but with lower enthalpies of formation.

\newpage
\section{\texorpdfstring{$\textbf{\ce{Nb2S}}$}{Nb2S}}
\begin{figure}[ht]
	\includegraphics[width=0.6\columnwidth]{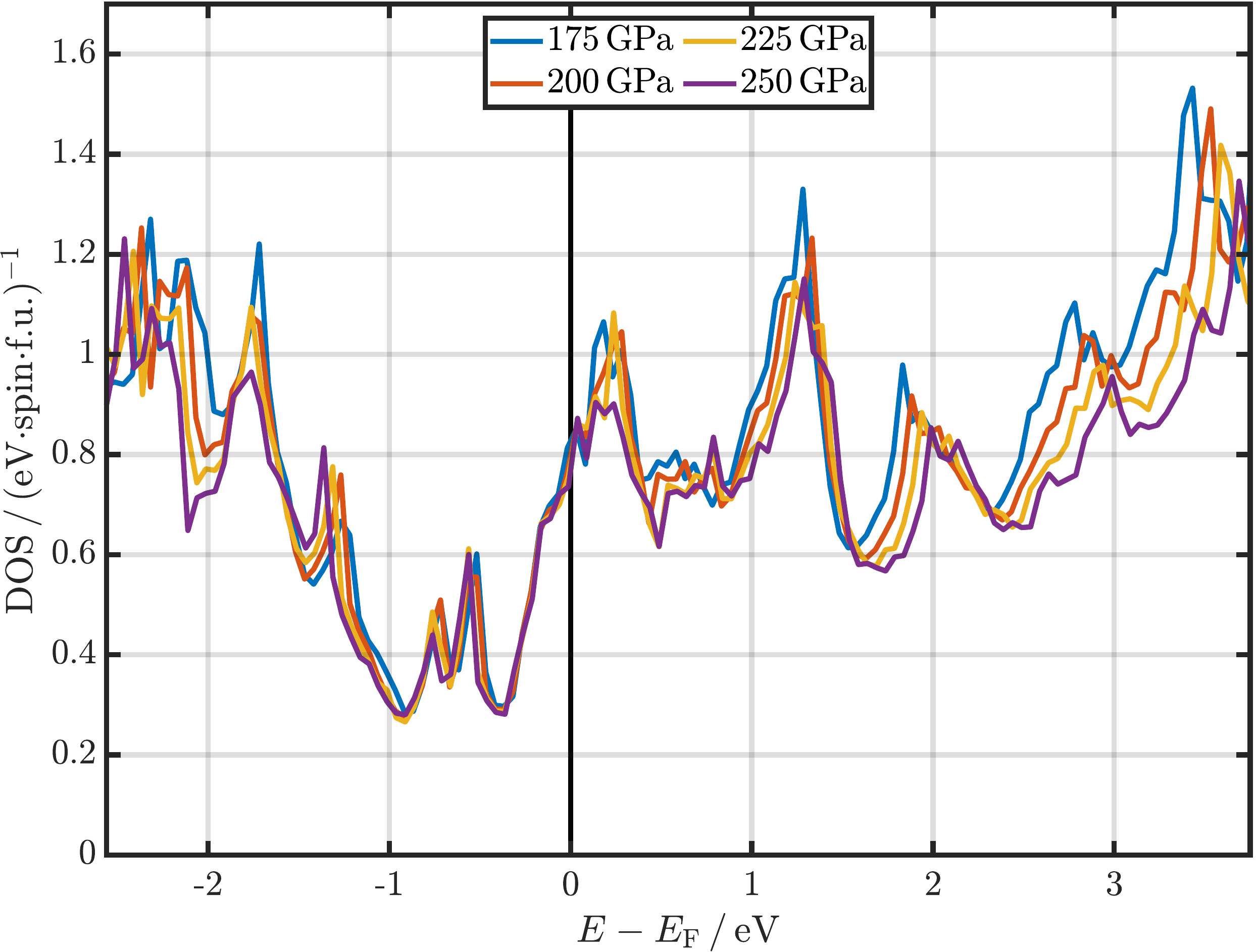}
	\caption{
		Electronic DOS of \ce{Nb2S} around the Fermi Energy $E_{\mathrm{F}}$ for different pressures.
	} 
	\label{fig:Nb2S_add_DOS}
\end{figure}

Figure~\ref{fig:Nb2S_add_DOS} shows the electronic DOS in \ce{Nb2S} for pressures from \SIrange{175}{250}{\giga\pascal}, emphasizing that the slope around the Fermi energy $E_{\mathrm{F}}$ does not change significantly with pressure.

\begin{figure}[ht]
	\includegraphics[width=0.6\columnwidth]{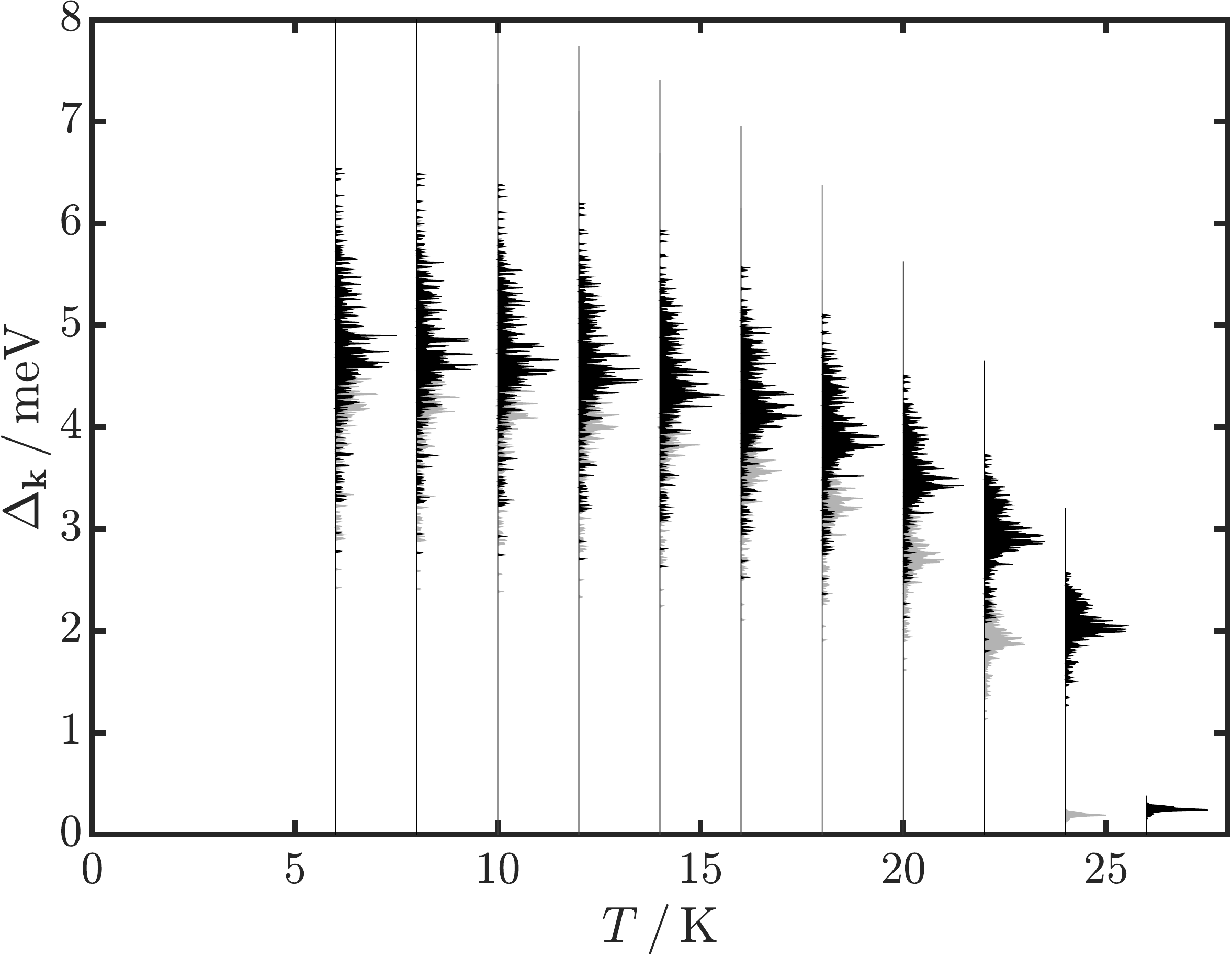}
	\caption{
		Energy distribution of the superconducting gap $\Delta_{\bf k}$ of \ce{Nb2S} at \SI{225}{\giga\pascal} as a function of temperature, calculated within the anisotropic Migdal-Eliashberg theory as implemented in the \textsc{EPW} code~\cite{ponce_epw:_2016}, using $\mu^* = 0.1$ (black) and $\mu^* = 0.14$ (grey).
	} 
	\label{fig:sc_gap}
\end{figure}

Figure~\ref{fig:sc_gap} shows the results from fully anisotropic Migdal-Eliashberg calculations for \ce{Nb2S} at \SI{225}{\giga\pascal}, revealing a single isotropic superconducting gap closing at $\sim$\SI{25}{\kelvin}. As reported before, the fully anisotropic, mode and wave vector resolved Migdal-Eliashberg calculations lead, in general, to higher $T_c$'s~\cite{margine_anisotropic_2013,heil2017NbS2,Heil_superconductivity_2019} than the averaged, isotropic Allen-Dynes McMillan formula~\cite{allen_transition_1975}.

\vfill
\newpage

\begin{figure}[ht]
	\includegraphics[width=0.8\columnwidth]{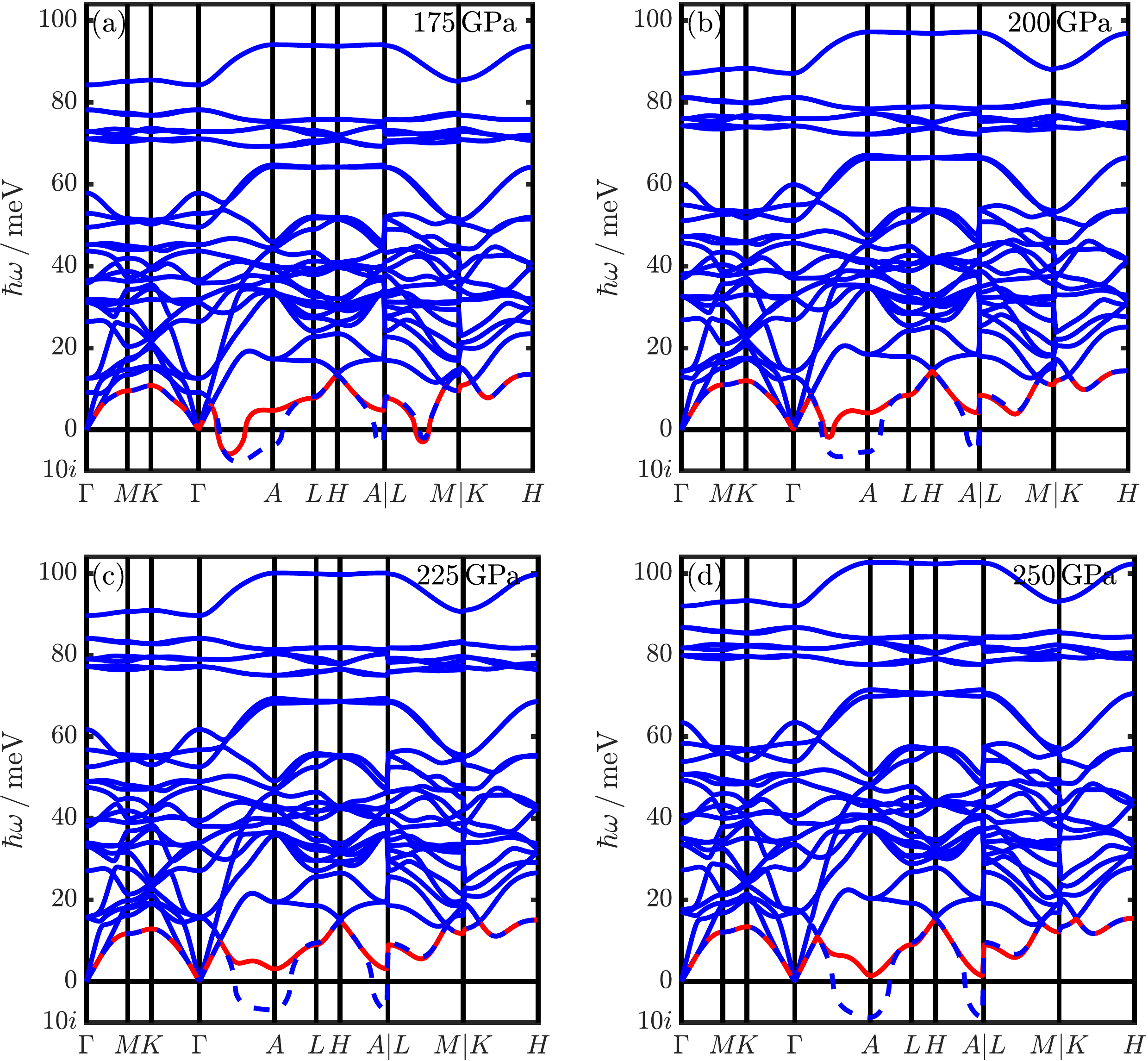}
	\caption{
		Phonon dispersions for \ce{Nb2S} for pressures (a) \SI{175}{\giga\pascal}, (b) \SI{200}{\giga\pascal}, (b) \SI{225}{\giga\pascal}, (d) \SI{250}{\giga\pascal}.
	} 
	\label{fig:Nb2S_add_ph}
\end{figure}

Figure~\ref{fig:Nb2S_add_ph} shows the full phonon dispersion curves for \ce{Nb2S} in the studied pressure range, including the anharmonic corrections for $\mathbf{q}_{1,2}$, where the opposed behaviour of the modes at $A$ and between $\Gamma-A$ can be appreciated.

\clearpage
\section{Electron-phonon properties of studied phases}
 The electron-phonon and superconducting properties of considered, metallic phases are summarized in Tab.~\ref{tab:elph}.
 
\begin{table}[h]
 	\caption{
 		Electron-phonon and superconducting properties of considered, metallic phases. The asterisk * indicates phases with one or more imaginary frequencies in the harmonic approximation, who's contribution to the $ep$ coupling has been set to zero, the dagger symbol $^\dagger$ indicates metastability, and $N_{\mathrm{F}}$ is the DOS per spin and per formula unit at the Fermi level. The reported values were obtained with a broadening value of \SI{0.025}{Ry}.
 	}
\begin{tabular}{lcccccc}
	& $p\,/\,\mathrm{GPa}$ & space group & $\lambda$ & $\omega_{\mathrm{log}}\,/\,\mathrm{meV}$ & $N_{\mathrm{F}}\,/\,\mathrm{eV}^{-1}$ & $T_c\,/\,\mathrm{K}$ \\ 
	\hline 
	%\ce{S} & 0 & 70 & --  & --  & --  & --  \\ 
	%\ce{S} & 10 & 148 & --  & --  & --  & --  \\
	%\ce{S} & 25 & 142 & --  & --  & --  & --  \\
	%\ce{S} & 50 & 142 & --  & --  & --  & --  \\ 
	\ce{S} & 200 & 166 & 0.68 & 39.5 & 0.14 & 14.7 \\
	%\ce{NbS3} a & 0 & 11 & --  & --  & --  & --  \\
	\ce{NbS3} b & 50 & 11 & 0.86 & 21 & 0.46 & 13.2 \\ 
	2H-$\ce{NbS2}^*$ & 0 & 194 & 1.43 & 6.5 & 1.36 & 8.2 \\ 
	2H-\ce{NbS2} (anharm., Ref.~\onlinecite{heil2017NbS2}) & 0 & 194 & 1.49 & 12.6 & 1.36 & 11.5 \\
	$\ce{NbS2}^*$ & 50 & 139 & 0.72 & 21.6 & 0.56 & 9.5 \\
	$\ce{NbS2}^*$ & 100 & 139 & 0.54 & 30.8 & 0.5 & 5.7 \\
	\ce{NbS2} & 200 & 139 & 0.5 & 38.1 & 0.47 & 5.6 \\
	%\ce{Nb3S5} & 0 & 1 & --  & --  & --  & --  \\ 
	\ce{Nb2S3} & 0 & 160 & 0.21 & 23.3 & 0.61 & 0 \\ 
	\ce{Nb2S3} & 100 & 139 & 0.39 & 34.1 & 0.66 & 1.5 \\
	\ce{Nb2S3} & 200 & 139 & 0.35 & 40.2 & 0.63 & 0.8 \\
	%\ce{Nb3S4} & 0 & 176 & --  & --  & --  & --  \\ 
	$\ce{NbS}^*$ & 50 & 194 & 0.63 & 25.1 & 0.56 & 7.8 \\ 
	\ce{NbS} & 100 & 221 & 0.56 & 31.9 & 0.68 & 6.8 \\
	\ce{NbS} & 150 & 221 & 0.49 & 36.2 & 0.58 & 4.9 \\
	\ce{NbS} & 200 & 221 & 0.46 & 39.4 & 0.51 & 3.8 \\
	$\ce{Nb2S}^{*\dagger}$ & 0 & 42 & 0.37 & 19.6 & 0.53 & 0.6 \\ 
	$\ce{Nb2S}^\dagger$ & 25 & 11 & 0.37 & 25.8 & 0.62 & 0.8 \\
	$\ce{Nb2S}^*$ (harm.) & 225 & 189 & 1.38 & 11.8 & 0.76 & 14.4 \\
	%\ce{Nb14S5} & 0 & 62 & --  & --  & --  & --  \\ 
	\ce{Nb3S} & 200 & 223 & 0.4 & 37 & 0.72 & 1.7 \\ 
	\ce{Nb} & 200 & 229 & 0.54 & 32.4 & 0.33 & 5.9 \\ 
	\hline
\end{tabular}

\label{tab:elph}
\end{table}

%\vfill
%\clearpage
\section{Crystal structure details}
In the following Tables~\ref{tab:S_0GPa}-\ref{tab:Nb_200GPa} we provide crystal structure details for all considered phases.

\begin{table}[h]
	\caption{Structural details for \ce{S} at \SI{0}{\giga\pascal}} \label{tab:S_0GPa}
	\begin{tabular}{lcccccccccc}
		& space group & $p\,/\,\mathrm{GPa}$ &  $a\,/$\AA & $b/a$ & $c/a$ & atom & site & $x$ & $y$ & $z$ \\ 
		\hline
		\ce{S} & 70 & 0 & 24.497 & 0.520 & 0.422 & S & $32h$ & $+0.051$ & $+0.045$ & $+0.144$ \\ 
		&  &  &  &  &  & S & $32h$ & $+0.003$ & $+0.485$ & $+0.204$ \\ 
		&  &  &  &  &  & S & $32h$ & $-0.424$ & $+0.036$ & $+0.284$ \\ 
		&  &  &  &  &  & S & $32h$ & $+0.131$ & $+0.410$ & $+0.284$ \\ 
		\hline 
	\end{tabular}
\end{table}

%\FloatBarrier

\begin{table}[h]
	\caption{Structural details for \ce{S} at \SI{10}{\giga\pascal}}
	\begin{tabular}{lcccccccccc}
		& space group & $p\,/\,\mathrm{GPa}$ &  $a\,/$\AA & $c/a$ & $\gamma\,/\,^\circ$ & atom & site & $x$ & $y$ & $z$ \\ 
		\hline
		\ce{S} & 148 & 10 & 9.952 & 0.362 & 120.00 & S & $18f$ & $+0.171$ & $+0.462$ & $+0.035$ \\ 
		\hline 
	\end{tabular}
\end{table}

%\FloatBarrier

\begin{table}[h]
	\caption{Structural details for \ce{S} at \SI{50}{\giga\pascal}}
	\begin{tabular}{lccccccccc}
		& space group & $p\,/\,\mathrm{GPa}$ &  $a\,/$\AA & $c/a$ & atom & site & $x$ & $y$ & $z$ \\ 
		\hline
		\ce{S} & 142 & 50 & 7.940 & 0.398 & S & $16f$ & $-0.369$ & $-0.119$ & $+0.125$ \\ 
		\hline 
	\end{tabular}
\end{table}

%\FloatBarrier

\begin{table}[h]
	\caption{Structural details for \ce{S} at \SI{200}{\giga\pascal}}
	\begin{tabular}{lcccccccccc}
		& space group & $p\,/\,\mathrm{GPa}$ &  $a\,/$\AA & $c/a$ & $\gamma\,/\,^\circ$ & atom & site & $x$ & $y$ & $z$ \\ 
		\hline
		\ce{S} & 166 & 200 & 3.323 & 0.782 & 120.00 & S & $3a$ & $+0.000$ & $+0.000$ & $+0.000$ \\ 
		\hline 
	\end{tabular}
\end{table}

%\FloatBarrier

\begin{table}[h]
	\caption{Structural details for \ce{NbS3} a at \SI{0}{\giga\pascal}}
	\begin{tabular}{lccccccccccc}
		& space group & $p\,/\,\mathrm{GPa}$ &  $a\,/$\AA & $b/a$ & $c/a$ & $\beta\,/\,^\circ$ & atom & site & $x$ & $y$ & $z$ \\ 
		\hline
		\ce{NbS3} a & 11 & 0 & 14.709 & 0.228 & 0.651 & 107.83 & Nb & $2e$ & $-0.119$ & $+0.250$ & $-0.472$ \\ 
		&  &  &  &  &  &  & Nb & $2e$ & $+0.205$ & $+0.250$ & $-0.108$ \\ 
		&  &  &  &  &  &  & Nb & $2e$ & $-0.460$ & $+0.250$ & $-0.163$ \\ 
		&  &  &  &  &  &  & S & $2e$ & $+0.098$ & $+0.250$ & $+0.266$ \\ 
		&  &  &  &  &  &  & S & $2e$ & $+0.236$ & $+0.250$ & $+0.430$ \\ 
		&  &  &  &  &  &  & S & $2e$ & $+0.431$ & $+0.250$ & $+0.342$ \\ 
		&  &  &  &  &  &  & S & $2e$ & $-0.428$ & $+0.250$ & $+0.329$ \\ 
		&  &  &  &  &  &  & S & $2e$ & $+0.066$ & $+0.250$ & $-0.373$ \\ 
		&  &  &  &  &  &  & S & $2e$ & $-0.085$ & $+0.250$ & $+0.085$ \\ 
		&  &  &  &  &  &  & S & $2e$ & $-0.276$ & $+0.250$ & $-0.096$ \\ 
		&  &  &  &  &  &  & S & $2e$ & $-0.249$ & $+0.250$ & $+0.274$ \\ 
		&  &  &  &  &  &  & S & $2e$ & $+0.405$ & $+0.250$ & $-0.039$ \\ 
		\hline 
	\end{tabular}
\end{table}

%\FloatBarrier

\begin{table}[h]
	\caption{Structural details for \ce{NbS3} b at \SI{50}{\giga\pascal}}
	\begin{tabular}{lccccccccccc}
		& space group & $p\,/\,\mathrm{GPa}$ &  $a\,/$\AA & $b/a$ & $c/a$ & $\beta\,/\,^\circ$ & atom & site & $x$ & $y$ & $z$ \\ 
		\hline
		\ce{NbS3} b & 11 & 50 & 4.551 & 0.670 & 1.691 & 98.28 & Nb & $2e$ & $+0.208$ & $+0.250$ & $-0.179$ \\ 
		&  &  &  &  &  &  & S & $2e$ & $-0.417$ & $+0.250$ & $+0.367$ \\ 
		&  &  &  &  &  &  & S & $2e$ & $-0.264$ & $+0.250$ & $-0.064$ \\ 
		&  &  &  &  &  &  & S & $2e$ & $+0.096$ & $+0.250$ & $+0.361$ \\ 
		\hline 
	\end{tabular}
\end{table}

%\FloatBarrier

\begin{table}[h]
	\caption{Structural details for 2H-\ce{NbS2}* at \SI{0}{\giga\pascal}}
	\begin{tabular}{lcccccccccc}
		& space group & $p\,/\,\mathrm{GPa}$ &  $a\,/$\AA & $c/a$ & $\gamma\,/\,^\circ$ & atom & site & $x$ & $y$ & $z$ \\ 
		\hline
		2H-\ce{NbS2}* & 194 & 0 & 3.327 & 3.576 & 120.00 & Nb & $2b$ & $+0.000$ & $+0.000$ & $+0.250$ \\ 
		&  &  &  &  &  & S & $4f$ & $+0.333$ & $+0.667$ & $+0.118$ \\ 
		\hline 
	\end{tabular}
\end{table}

%\FloatBarrier

\begin{table}[h]
	\caption{Structural details for \ce{NbS2} at \SI{200}{\giga\pascal}}
	\begin{tabular}{lccccccccc}
		& space group & $p\,/\,\mathrm{GPa}$ &  $a\,/$\AA & $c/a$ & atom & site & $x$ & $y$ & $z$ \\ 
		\hline
		\ce{NbS2} & 139 & 200 & 2.817 & 2.509 & Nb & $2a$ & $+0.000$ & $+0.000$ & $+0.000$ \\ 
		&  &  &  &  & S & $4e$ & $+0.000$ & $+0.000$ & $+0.332$ \\ 
		\hline 
	\end{tabular}
\end{table}

%\FloatBarrier

\begin{table}[h]
	\caption{Structural details for \ce{Nb3S5} at \SI{0}{\giga\pascal}}
	\begin{tabular}{lccccccccccccc}
		& space group & $p\,/\,\mathrm{GPa}$ &  $a\,/$\AA & $b/a$ & $c/a$ & $\alpha\,/\,^\circ$ & $\beta\,/\,^\circ$ & $\gamma\,/\,^\circ$ & atom & site & $x$ & $y$ & $z$ \\ 
		\hline
		\ce{Nb3S5} & 1 & 0 & 5.812 & 1.219 & 1.351 & 72.34 & 97.21 & 74.23 & Nb & $1a$ & $+0.330$ & $-0.474$ & $+0.018$ \\ 
		&  &  &  &  &  &  &  &  & Nb & $1a$ & $-0.001$ & $-0.000$ & $-0.000$ \\ 
		&  &  &  &  &  &  &  &  & Nb & $1a$ & $+0.405$ & $+0.210$ & $-0.203$ \\ 
		&  &  &  &  &  &  &  &  & Nb & $1a$ & $-0.193$ & $+0.403$ & $-0.395$ \\ 
		&  &  &  &  &  &  &  &  & Nb & $1a$ & $+0.206$ & $-0.397$ & $+0.396$ \\ 
		&  &  &  &  &  &  &  &  & Nb & $1a$ & $-0.400$ & $-0.201$ & $+0.202$ \\ 
		&  &  &  &  &  &  &  &  & S & $1a$ & $-0.228$ & $+0.048$ & $-0.305$ \\ 
		&  &  &  &  &  &  &  &  & S & $1a$ & $-0.052$ & $-0.356$ & $+0.094$ \\ 
		&  &  &  &  &  &  &  &  & S & $1a$ & $+0.172$ & $+0.247$ & $+0.497$ \\ 
		&  &  &  &  &  &  &  &  & S & $1a$ & $+0.370$ & $-0.139$ & $-0.118$ \\ 
		&  &  &  &  &  &  &  &  & S & $1a$ & $-0.411$ & $+0.427$ & $+0.312$ \\ 
		&  &  &  &  &  &  &  &  & S & $1a$ & $+0.100$ & $-0.431$ & $-0.297$ \\ 
		&  &  &  &  &  &  &  &  & S & $1a$ & $+0.299$ & $+0.160$ & $+0.109$ \\ 
		&  &  &  &  &  &  &  &  & S & $1a$ & $+0.500$ & $-0.237$ & $-0.496$ \\ 
		&  &  &  &  &  &  &  &  & S & $1a$ & $-0.295$ & $+0.364$ & $-0.091$ \\ 
		&  &  &  &  &  &  &  &  & S & $4e$ & $-0.098$ & $-0.041$ & $+0.304$ \\ 
		\hline 
	\end{tabular}
\end{table}

%\FloatBarrier

\begin{table}[h]
	\caption{Structural details for \ce{Nb2S3} at \SI{0}{\giga\pascal}}
	\begin{tabular}{lcccccccccc}
		& space group & $p\,/\,\mathrm{GPa}$ &  $a\,/$\AA & $c/a$ & $\gamma\,/\,^\circ$ & atom & site & $x$ & $y$ & $z$ \\ 
		\hline
		\ce{Nb2S3} & 160 & 0 & 3.294 & 8.576 & 120.00 & Nb & $3a$ & $+0.000$ & $+0.000$ & $-0.392$ \\ 
		&  &  &  &  &  & Nb & $3a$ & $+0.000$ & $+0.000$ & $-0.274$ \\ 
		&  &  &  &  &  & S & $3a$ & $+0.000$ & $+0.000$ & $+0.115$ \\ 
		&  &  &  &  &  & S & $3a$ & $+0.000$ & $+0.000$ & $-0.115$ \\ 
		&  &  &  &  &  & S & $6h$ & $+0.000$ & $+0.000$ & $+0.000$ \\ 
		\hline 
	\end{tabular}
\end{table}

%\FloatBarrier

\begin{table}[h]
	\caption{Structural details for \ce{Nb2S3} at \SI{200}{\giga\pascal}}
	\begin{tabular}{lccccccccc}
		& space group & $p\,/\,\mathrm{GPa}$ &  $a\,/$\AA & $c/a$ & atom & site & $x$ & $y$ & $z$ \\ 
		\hline
		\ce{Nb2S3} & 139 & 200 & 2.762 & 4.498 & Nb & $2a$ & $+0.000$ & $+0.000$ & $-0.391$ \\ 
		&  &  &  &  & S & $4e$ & $+0.000$ & $+0.000$ & $+0.000$ \\ 
		&  &  &  &  & S & $4e$ & $+0.000$ & $+0.000$ & $-0.200$ \\ 
		\hline 
	\end{tabular}
\end{table}

%\FloatBarrier

\begin{table}[h]
	\caption{Structural details for \ce{Nb3S4} at \SI{0}{\giga\pascal}}
	\begin{tabular}{lcccccccccc}
		& space group & $p\,/\,\mathrm{GPa}$ &  $a\,/$\AA & $c/a$ & $\gamma\,/\,^\circ$ & atom & site & $x$ & $y$ & $z$ \\ 
		\hline
		\ce{Nb3S4} & 176 & 0 & 9.589 & 0.352 & 120.00 & Nb & $6h$ & $-0.114$ & $-0.486$ & $+0.250$ \\ 
		&  &  &  &  &  & S & $2c$ & $-0.286$ & $-0.344$ & $+0.250$ \\ 
		&  &  &  &  &  & S & $1b$ & $+0.333$ & $+0.667$ & $+0.250$ \\ 
		\hline 
	\end{tabular}
\end{table}

%\FloatBarrier

\begin{table}[h]
	\caption{Structural details for \ce{NbS}* at \SI{50}{\giga\pascal}}
	\begin{tabular}{lcccccccccc}
		& space group & $p\,/\,\mathrm{GPa}$ &  $a\,/$\AA & $c/a$ & $\gamma\,/\,^\circ$ & atom & site & $x$ & $y$ & $z$ \\ 
		\hline
		\ce{NbS}* & 194 & 50 & 3.060 & 2.098 & 120.00 & Nb & $2c$ & $+0.000$ & $+0.000$ & $+0.000$ \\ 
		&  &  &  &  &  & S & $3f$ & $+0.333$ & $+0.667$ & $+0.250$ \\ 
		\hline  
	\end{tabular}
\end{table}

%\FloatBarrier

\begin{table}[h]
	\caption{Structural details for \ce{NbS} at \SI{200}{\giga\pascal}}
	\begin{tabular}{lcccccccc}
		& space group & $p\,/\,\mathrm{GPa}$ &  $a\,/$\AA & atom & site & $x$ & $y$ & $z$ \\ 
		\hline
		\ce{NbS} & 221 & 200 & 2.683 & Nb & $1a$ & $+0.500$ & $+0.500$ & $+0.500$ \\ 
		&  &  &  & S & $1b$ & $+0.000$ & $+0.000$ & $+0.000$ \\ 
		\hline 
	\end{tabular}
\end{table}

%\FloatBarrier

\begin{table}[h]
	\caption{Structural details for $\ce{Nb2S}*^\dagger$ at \SI{0}{\giga\pascal}}
	\begin{tabular}{lcccccccccc}
		& space group & $p\,/\,\mathrm{GPa}$ &  $a\,/$\AA & $b/a$ & $c/a$ & atom & site & $x$ & $y$ & $z$ \\ 
		\hline
		$\ce{Nb2S}*^\dagger$ & 42 & 0 & 3.307 & 3.468 & 1.590 & Nb & $4a$ & $+0.000$ & $-0.349$ & $-0.158$ \\ 
		&  &  &  &  &  & S & $4c$ & $+0.000$ & $+0.000$ & $+0.000$ \\ 
		\hline 
	\end{tabular}
\end{table}

%\FloatBarrier

\begin{table}[h]
	\caption{Structural details for $\ce{Nb2S}^\dagger$ at \SI{25}{\giga\pascal}}
	\begin{tabular}{lccccccccccc}
		& space group & $p\,/\,\mathrm{GPa}$ &  $a\,/$\AA & $b/a$ & $c/a$ & $\beta\,/\,^\circ$ & atom & site & $x$ & $y$ & $z$ \\ 
		\hline
		$\ce{Nb2S}^\dagger$ & 11 & 25 & 4.681 & 0.667 & 1.375 & 109.77 & Nb & $2e$ & $-0.298$ & $+0.250$ & $-0.068$ \\ 
		&  &  &  &  &  &  & Nb & $2e$ & $-0.303$ & $+0.250$ & $+0.417$ \\ 
		&  &  &  &  &  &  & S & $8c$ & $+0.106$ & $+0.250$ & $+0.277$ \\ 
		\hline 
	\end{tabular}
\end{table}

\FloatBarrier

\begin{table}[h]
	\caption{Structural details for \ce{Nb2S}* (harm) at \SI{225}{\giga\pascal}}
	\begin{tabular}{lcccccccccc}
		& space group & $p\,/\,\mathrm{GPa}$ &  $a\,/$\AA & $c/a$ & $\gamma\,/\,^\circ$ & atom & site & $x$ & $y$ & $z$ \\ 
		\hline
		\ce{Nb2S}* (harm) & 189 & 225 & 6.018 & 0.478 & 120.00 & Nb & $3g$ & $-0.401$ & $+0.000$ & $+0.000$ \\ 
		&  &  &  &  &  & Nb & $1a$ & $+0.264$ & $+0.000$ & $+0.500$ \\ 
		&  &  &  &  &  & S & $2d$ & $+0.000$ & $+0.000$ & $+0.000$ \\ 
		&  &  &  &  &  & S & $2e$ & $+0.333$ & $+0.667$ & $+0.500$ \\ 
		\hline 
	\end{tabular}
\end{table}

%\FloatBarrier

\begin{table}[h]
	\caption{Structural details for \ce{Nb14S5} at \SI{0}{\giga\pascal}}
	\begin{tabular}{lcccccccccc}
		& space group & $p\,/\,\mathrm{GPa}$ &  $a\,/$\AA & $b/a$ & $c/a$ & atom & site & $x$ & $y$ & $z$ \\ 
		\hline
		\ce{Nb14S5} & 62 & 0 & 18.533 & 0.182 & 1.070 & Nb & $4c$ & $-0.467$ & $+0.250$ & $-0.456$ \\ 
		&  &  &  &  &  & Nb & $4c$ & $+0.491$ & $+0.250$ & $+0.226$ \\ 
		&  &  &  &  &  & Nb & $4c$ & $+0.052$ & $+0.250$ & $-0.385$ \\ 
		&  &  &  &  &  & Nb & $4c$ & $+0.398$ & $+0.250$ & $+0.365$ \\ 
		&  &  &  &  &  & Nb & $4c$ & $-0.138$ & $+0.250$ & $-0.019$ \\ 
		&  &  &  &  &  & Nb & $4c$ & $+0.050$ & $+0.250$ & $+0.449$ \\ 
		&  &  &  &  &  & Nb & $4c$ & $-0.171$ & $+0.250$ & $+0.348$ \\ 
		&  &  &  &  &  & Nb & $4c$ & $-0.255$ & $+0.250$ & $+0.204$ \\ 
		&  &  &  &  &  & Nb & $4c$ & $-0.430$ & $+0.250$ & $+0.379$ \\ 
		&  &  &  &  &  & Nb & $4c$ & $+0.299$ & $+0.250$ & $-0.322$ \\ 
		&  &  &  &  &  & Nb & $4c$ & $+0.141$ & $+0.250$ & $-0.242$ \\ 
		&  &  &  &  &  & Nb & $4c$ & $-0.281$ & $+0.250$ & $-0.419$ \\ 
		&  &  &  &  &  & Nb & $4c$ & $-0.118$ & $+0.250$ & $-0.334$ \\ 
		&  &  &  &  &  & Nb & $4c$ & $-0.298$ & $+0.250$ & $-0.035$ \\ 
		&  &  &  &  &  & S & $4c$ & $-0.097$ & $+0.250$ & $+0.467$ \\ 
		&  &  &  &  &  & S & $4c$ & $-0.379$ & $+0.250$ & $+0.261$ \\ 
		&  &  &  &  &  & S & $4c$ & $-0.290$ & $+0.250$ & $+0.420$ \\ 
		&  &  &  &  &  & S & $4c$ & $+0.178$ & $+0.250$ & $+0.411$ \\ 
		&  &  &  &  &  & S & $6c$ & $-0.486$ & $+0.250$ & $-0.322$ \\ 
		\hline 
	\end{tabular}
\end{table}

%\FloatBarrier

\begin{table}[h]
	\caption{Structural details for \ce{Nb3S} at \SI{200}{\giga\pascal}}
	\begin{tabular}{lcccccccc}
		& space group & $p\,/\,\mathrm{GPa}$ &  $a\,/$\AA & atom & site & $x$ & $y$ & $z$ \\ 
		\hline
		\ce{Nb3S} & 223 & 200 & 4.361 & Nb & $2a$ & $+0.250$ & $+0.000$ & $+0.500$ \\ 
		&  &  &  & S & $2a$ & $+0.000$ & $+0.000$ & $+0.000$ \\ 
		\hline 
	\end{tabular}
\end{table}

%\FloatBarrier

\begin{table}[h]
	\caption{Structural details for \ce{Nb} at \SI{200}{\giga\pascal}} \label{tab:Nb_200GPa}
	\begin{tabular}{lcccccccc}
		& space group & $p\,/\,\mathrm{GPa}$ &  $a\,/$\AA & atom & site & $x$ & $y$ & $z$ \\ 
		\hline
		\ce{Nb} & 229 & 200 & 2.816 & Nb & $6c$ & $+0.000$ & $+0.000$ & $+0.000$ \\ 
		\hline 
	\end{tabular}
\end{table}

\FloatBarrier
%\null
%\vfill
%\clearpage
\newpage
\section{Electronic and vibrational properties of studied phases}
In the following figures we report the electronic properties for all studied phases, where the partial DOS for \ce{Nb} $d$ orbitals is shown in green, \ce{Nb} $s$ in turquoise, \ce{Nb} $p$ in blue, and the \ce{S} $p$ orbitals in red and \ce{S} $s$ in ochre. We performed phonon calculations for phases for which we could not find corresponding data in literature.

\begin{figure}[ht]
	\includegraphics[width=0.77\columnwidth]{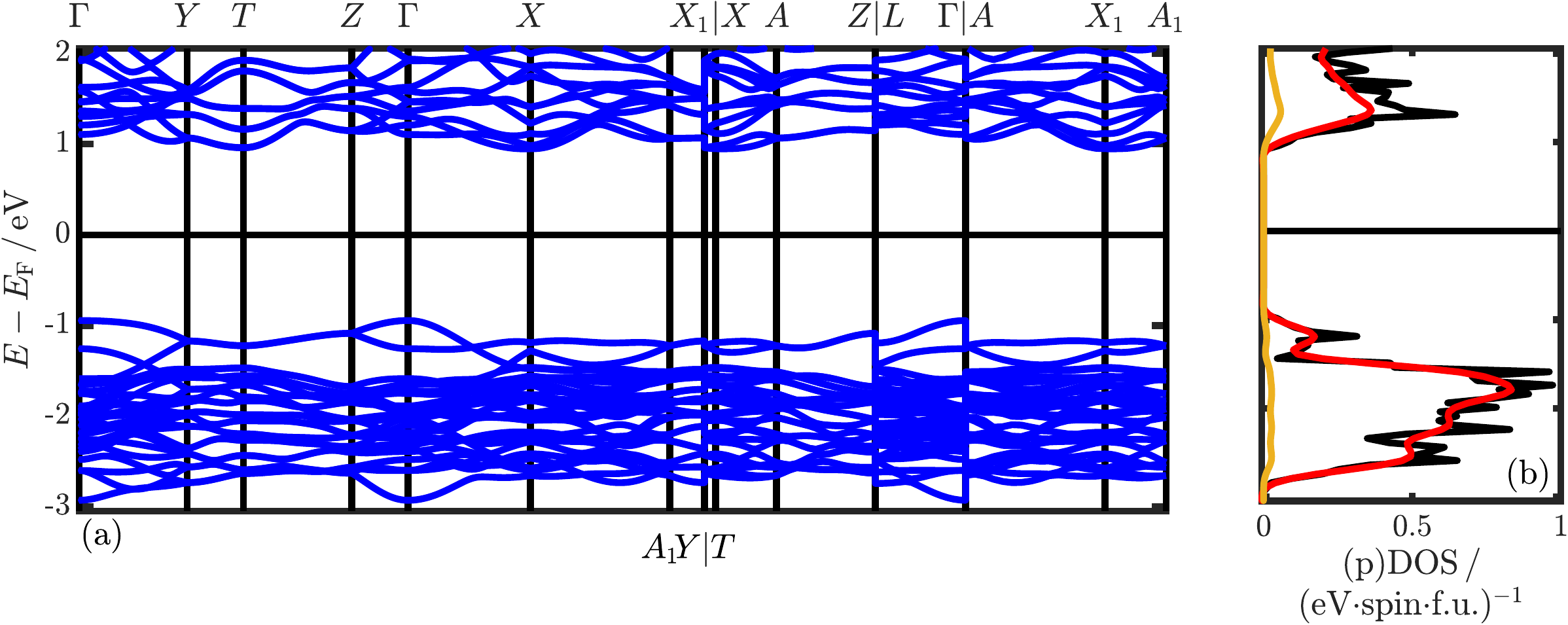}
	\caption{
		\ce{S} at \SI{0}{\giga\pascal}
	} 
	\label{fig:S_0}
\end{figure}

\begin{figure}[ht]
	\includegraphics[width=0.77\columnwidth]{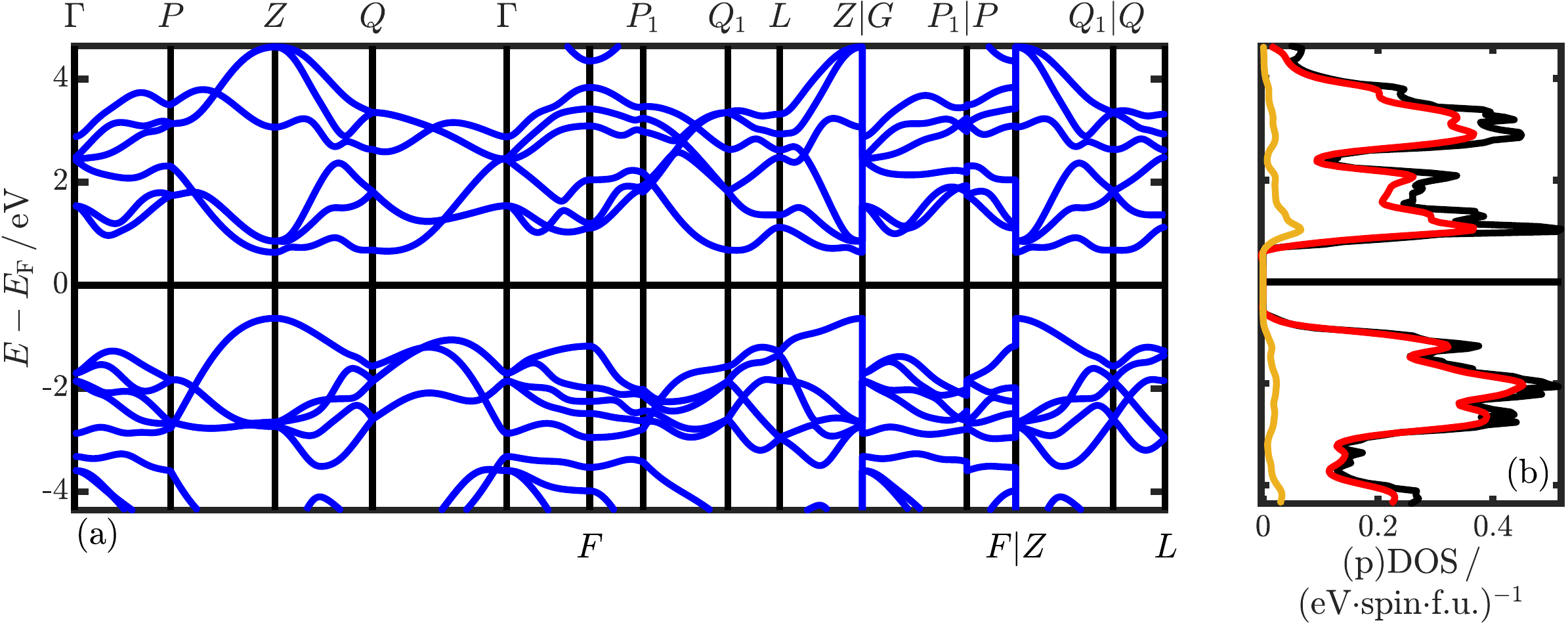}
	\caption{
		\ce{S} at \SI{10}{\giga\pascal}
	} 
	\label{fig:S_10}
\end{figure}

\begin{figure}[ht]
	\includegraphics[width=0.77\columnwidth]{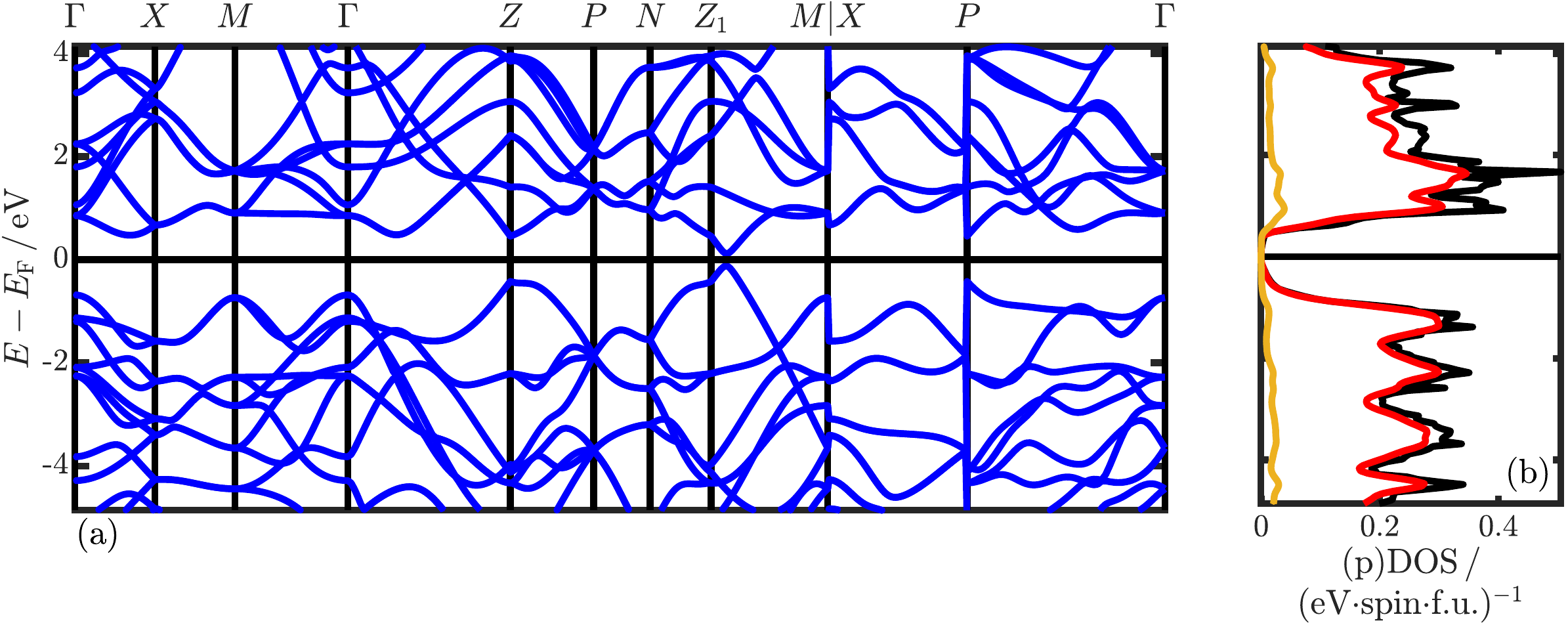}
	\caption{
		\ce{S} at \SI{25}{\giga\pascal} 
	} 
	\label{fig:S_25}
\end{figure}
 
\begin{figure}[ht]
	\includegraphics[width=0.77\columnwidth]{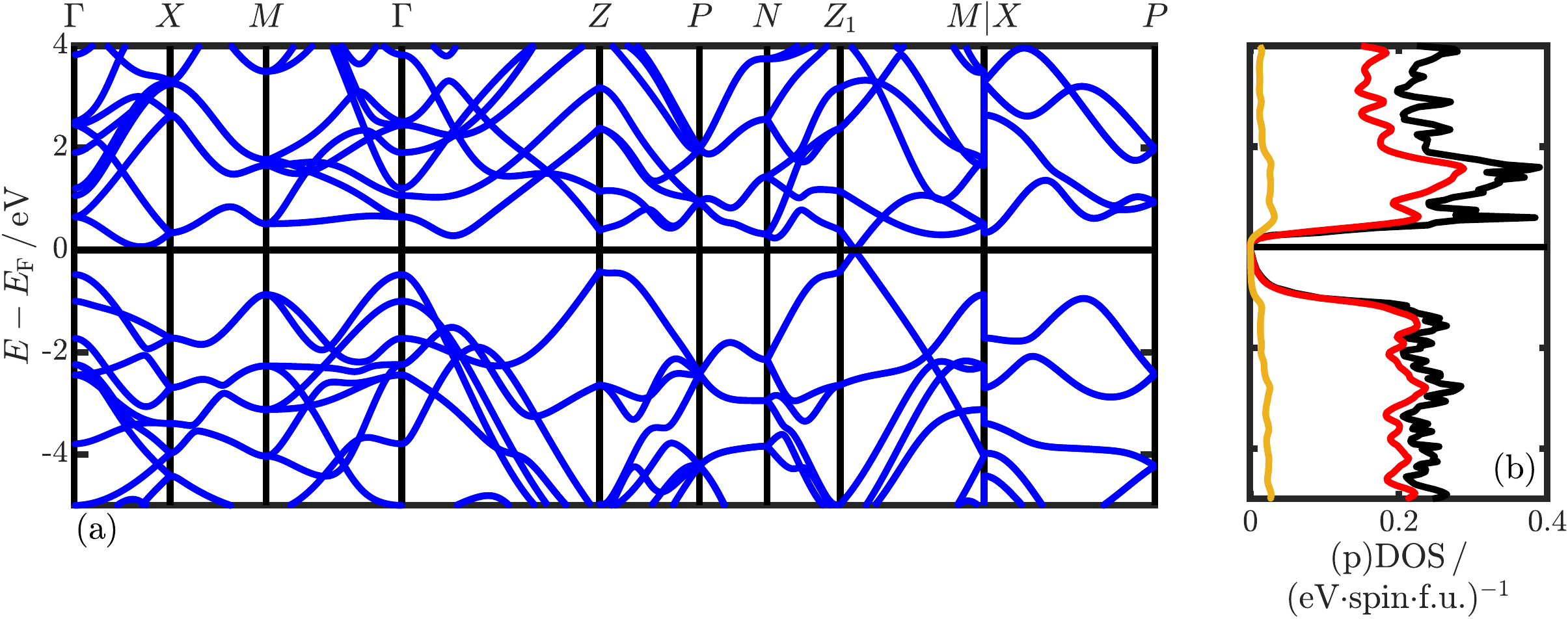}
	\caption{
		\ce{S} at \SI{50}{\giga\pascal}
	} 
	\label{fig:S_50}
\end{figure}
  
\begin{figure}[ht]
	\includegraphics[width=0.77\columnwidth]{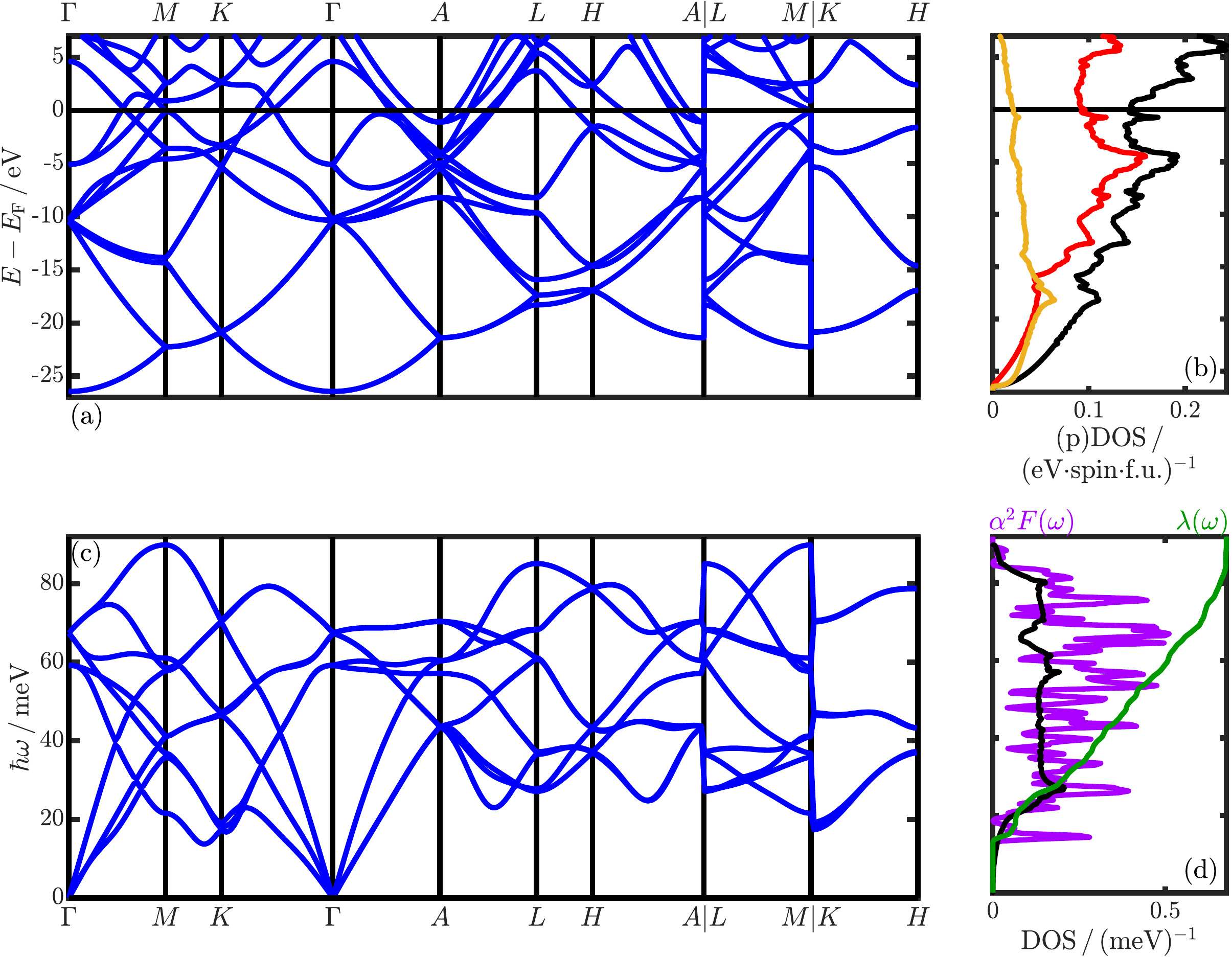}
	\caption{
		\ce{S} at \SI{200}{\giga\pascal}
	} 
	\label{fig:S_200}
\end{figure}

%\FloatBarrier
%\null
%\vfill
%\newpage

\begin{figure}[ht]
	\includegraphics[width=0.77\columnwidth]{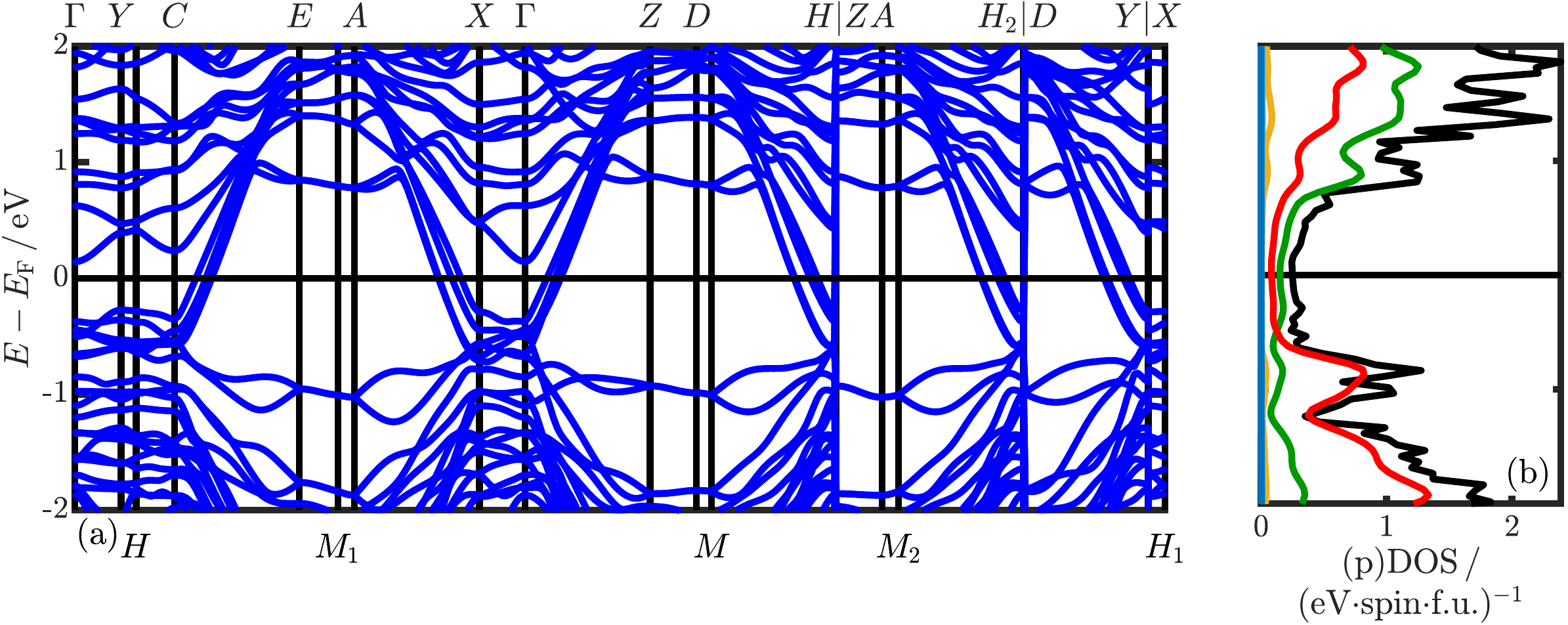}
	\caption{
		\ce{NbS3} at \SI{0}{\giga\pascal}
	} 
	\label{fig:NbS3_0}
\end{figure}
 
\begin{figure}[ht]
	\includegraphics[width=0.77\columnwidth]{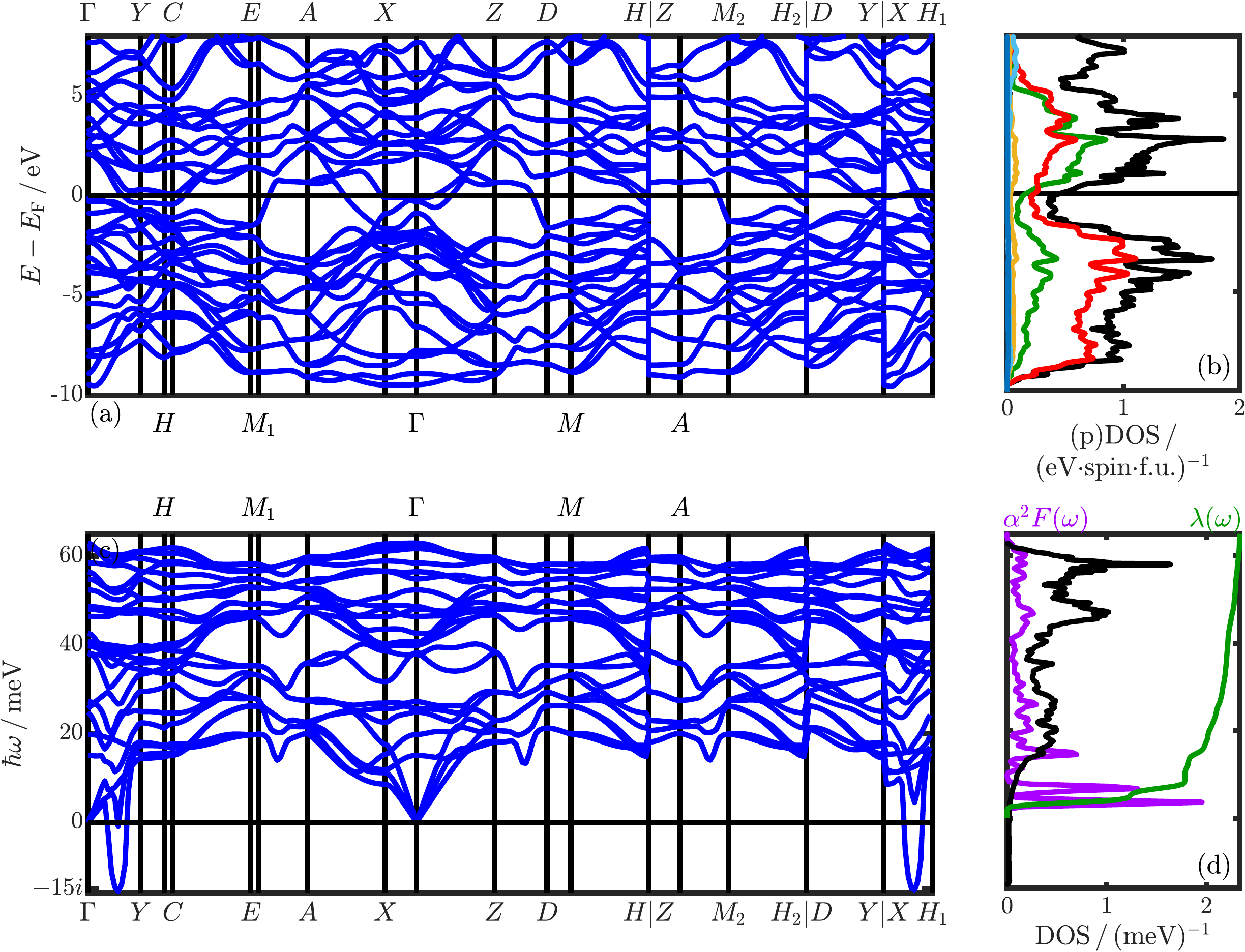}
	\caption{
		\ce{NbS3} at \SI{50}{\giga\pascal}
	} 
	\label{fig:NbS3_50}
\end{figure}
 
\begin{figure}[ht]
	\includegraphics[width=0.77\columnwidth]{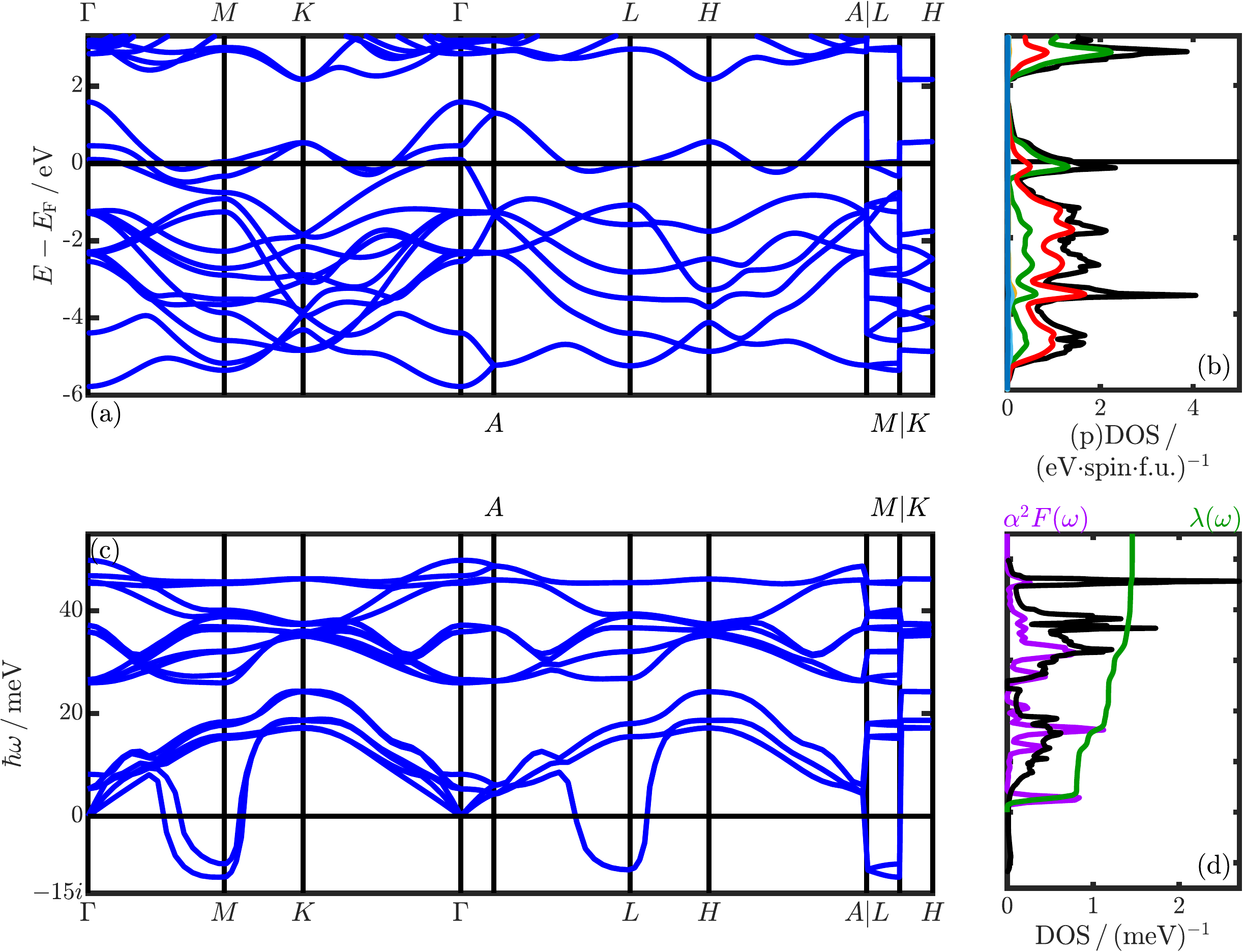}
	\caption{
		\ce{NbS2} at \SI{0}{\giga\pascal}
	} 
	\label{fig:NbS2_0}
\end{figure}
 
\begin{figure}[ht]
	\includegraphics[width=0.77\columnwidth]{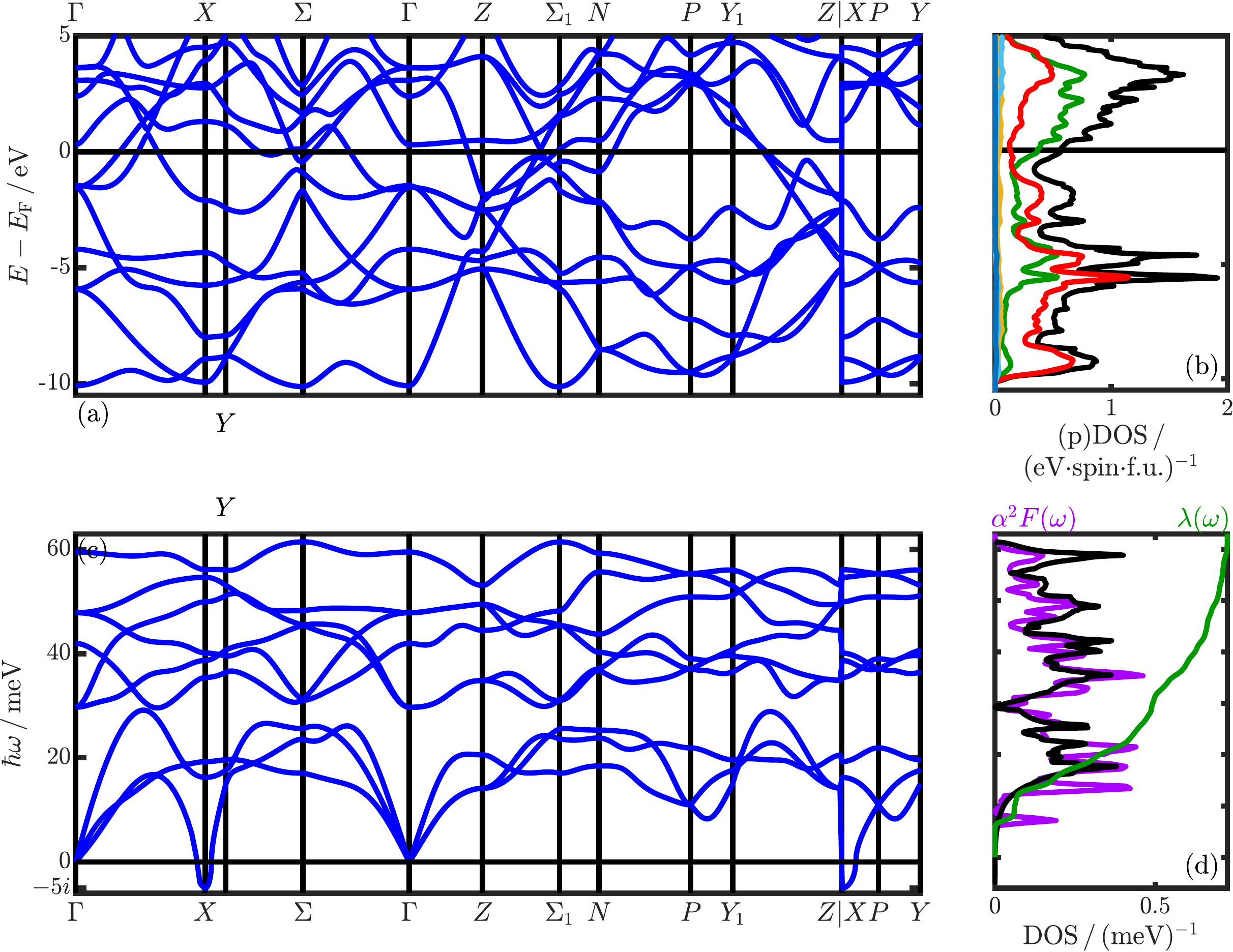}
	\caption{
		\ce{NbS2} at \SI{50}{\giga\pascal}
	} 
	\label{fig:NbS2_50}
\end{figure}
 
\begin{figure}[ht]
	\includegraphics[width=0.77\columnwidth]{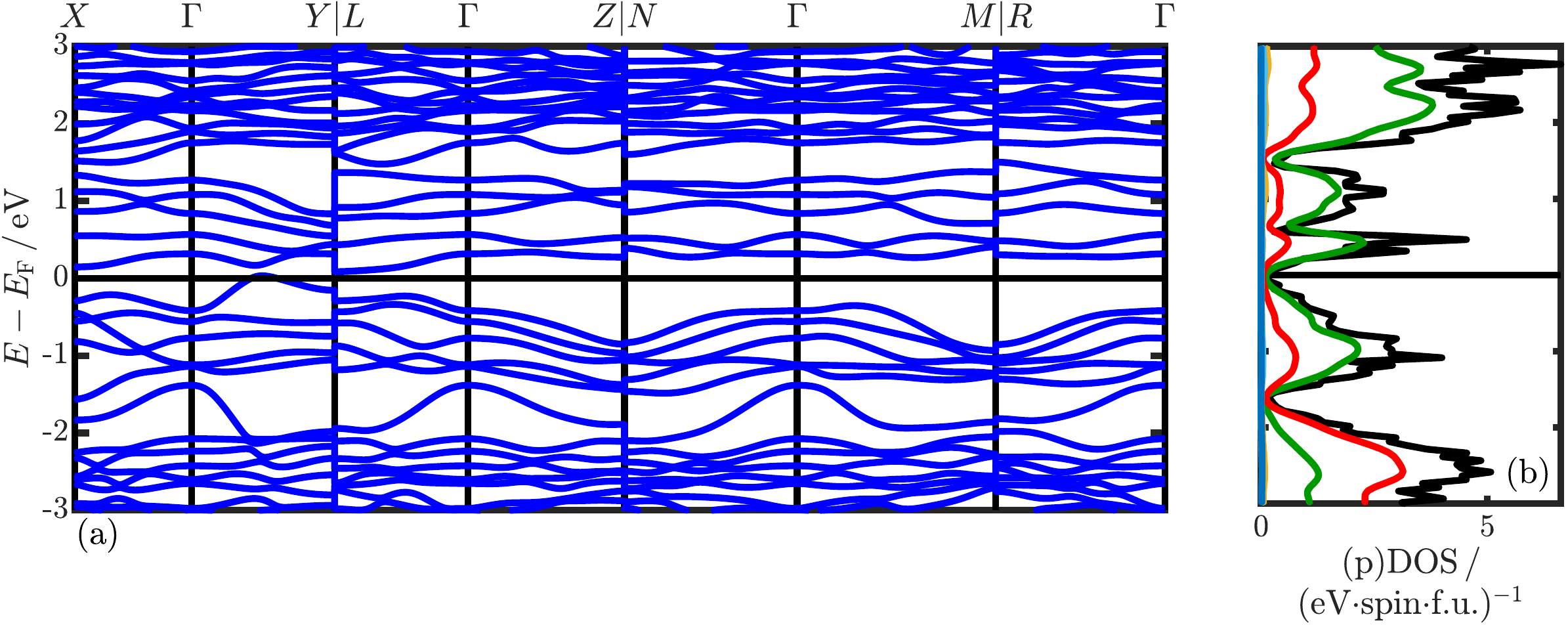}
	\caption{
		\ce{Nb3S5} at \SI{0}{\giga\pascal}
	} 
	\label{fig:Nb3S5_0}
\end{figure}
 
\begin{figure}[ht]
	\includegraphics[width=0.77\columnwidth]{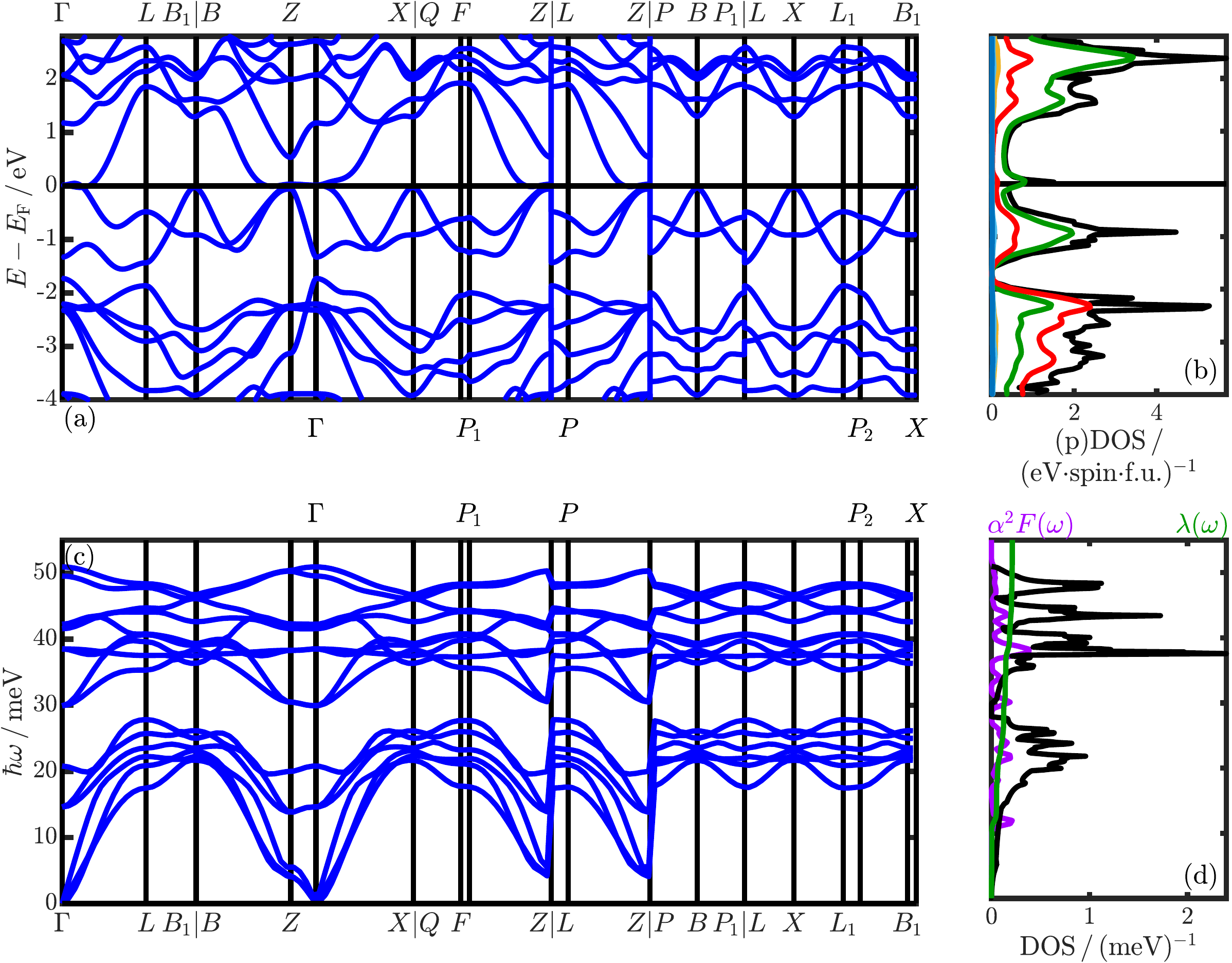}
	\caption{
		\ce{Nb2S3} at \SI{0}{\giga\pascal} 
	} 
	\label{fig:Nb2S3_0}
\end{figure}

%\FloatBarrier
%\null
%\vfill
%\newpage
 
\begin{figure}[ht]
	\includegraphics[width=0.77\columnwidth]{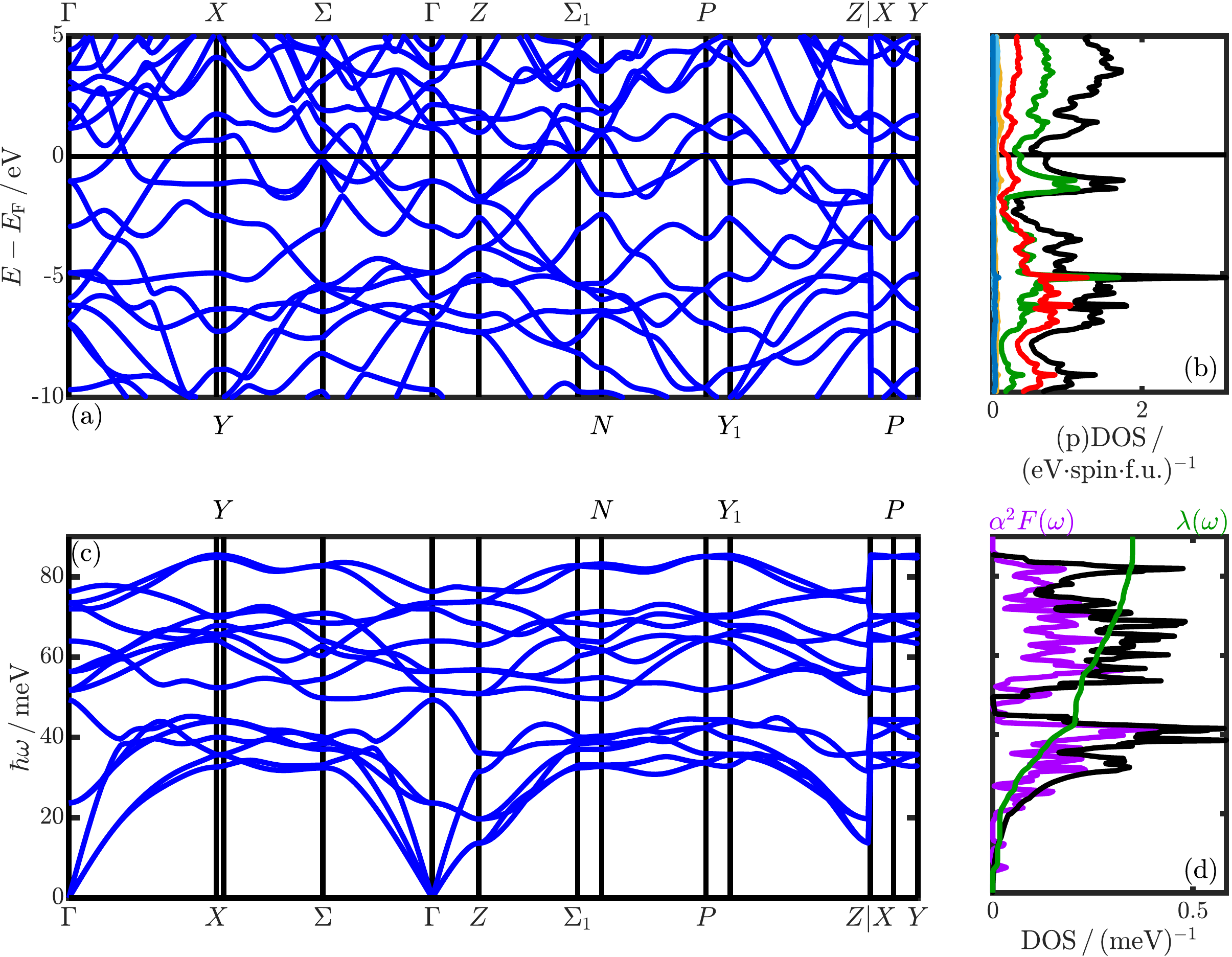}
	\caption{
		\ce{Nb2S3} at \SI{200}{\giga\pascal} 
	} 
	\label{fig:Nb2S3_200}
\end{figure}
 
\begin{figure}[ht]
	\includegraphics[width=0.77\columnwidth]{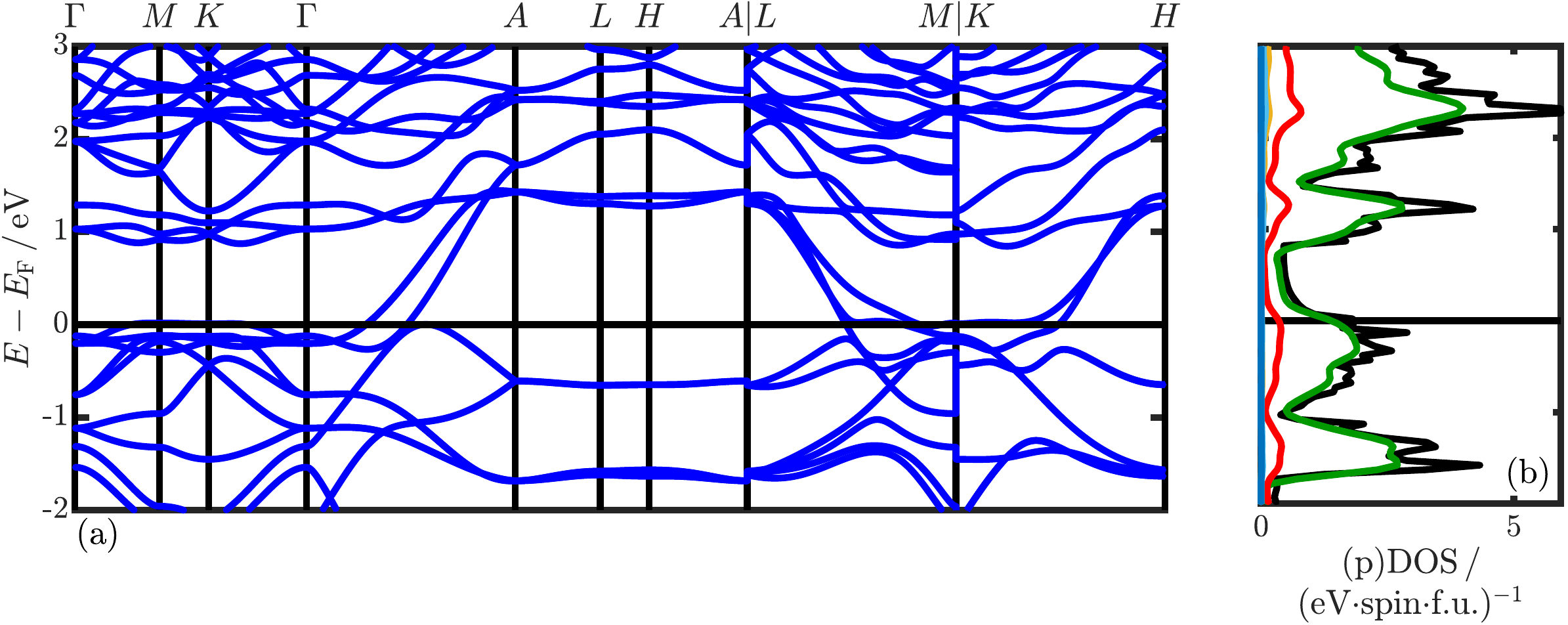}
	\caption{
		\ce{Nb3S4} at \SI{0}{\giga\pascal} 
	} 
	\label{fig:Nb3S4_0}
\end{figure}

%\FloatBarrier
%\null
%\vfill
%\newpage
 
\begin{figure}[ht]
	\includegraphics[width=0.77\columnwidth]{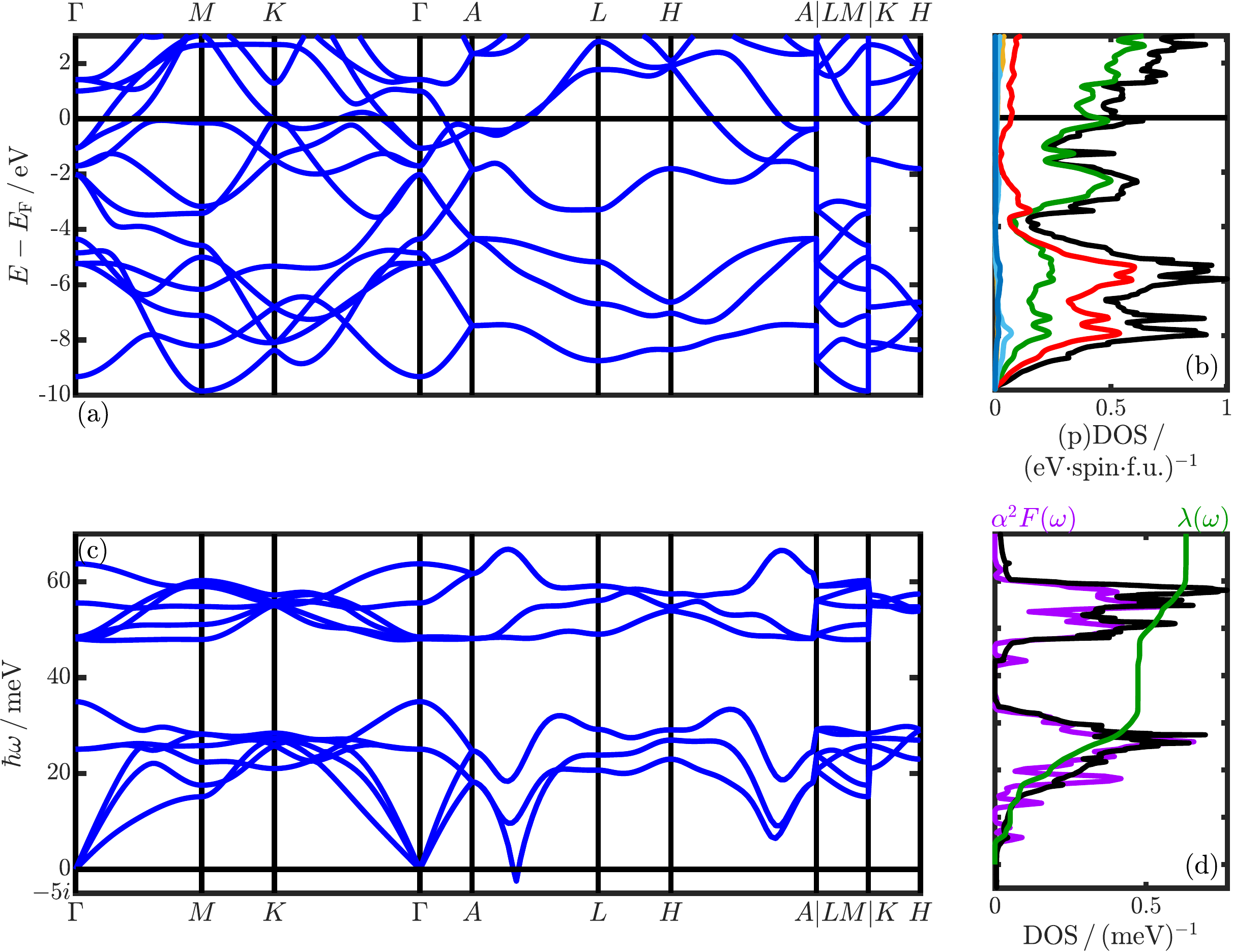}
	\caption{
		\ce{NbS} at \SI{50}{\giga\pascal}
	} 
	\label{fig:NbS_50}
\end{figure}
 
\begin{figure}[ht]
	\includegraphics[width=0.77\columnwidth]{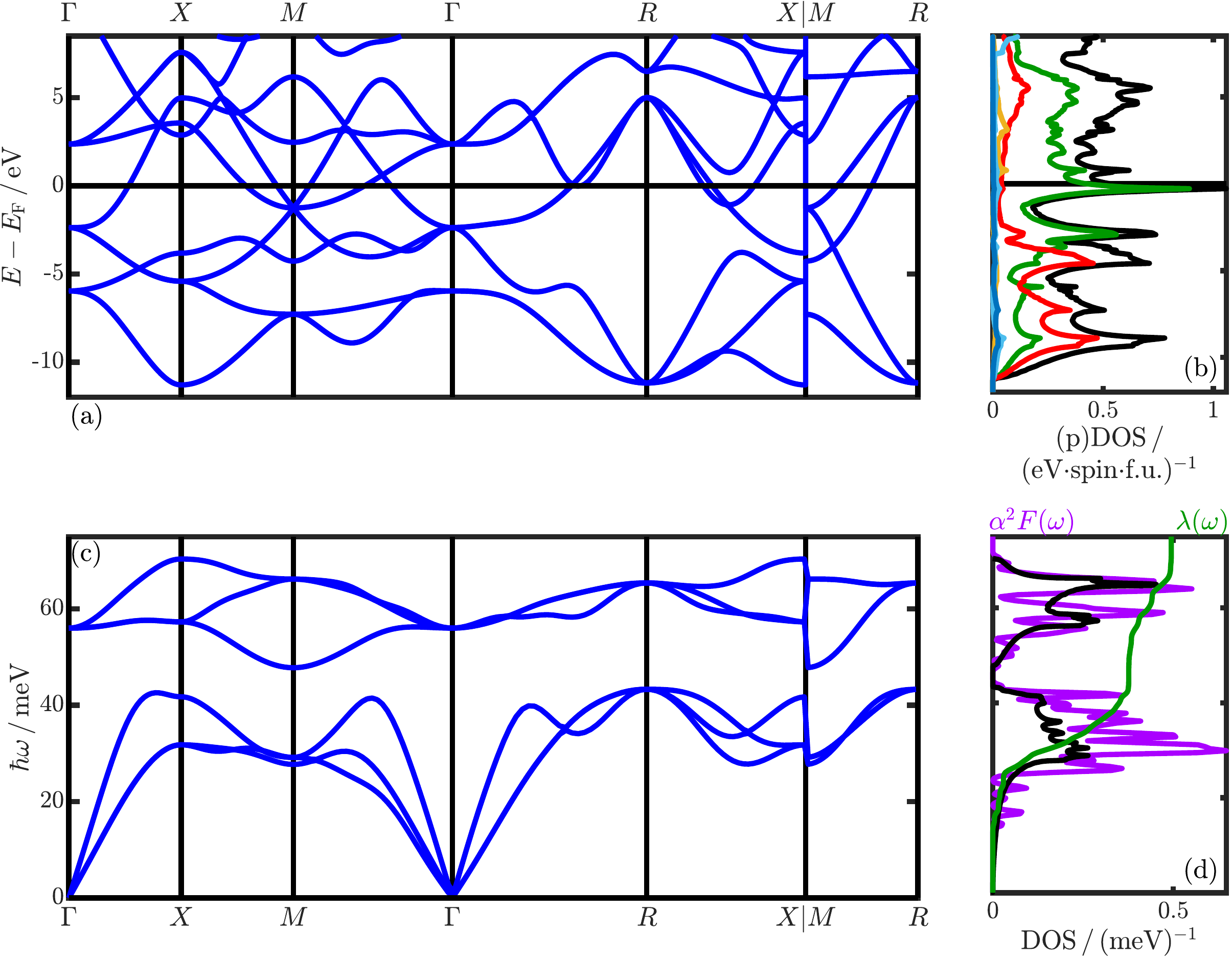}
	\caption{
		\ce{NbS} at \SI{150}{\giga\pascal}
	} 
	\label{fig:NbS_150}
\end{figure}
 
\begin{figure}[ht]
	\includegraphics[width=0.77\columnwidth]{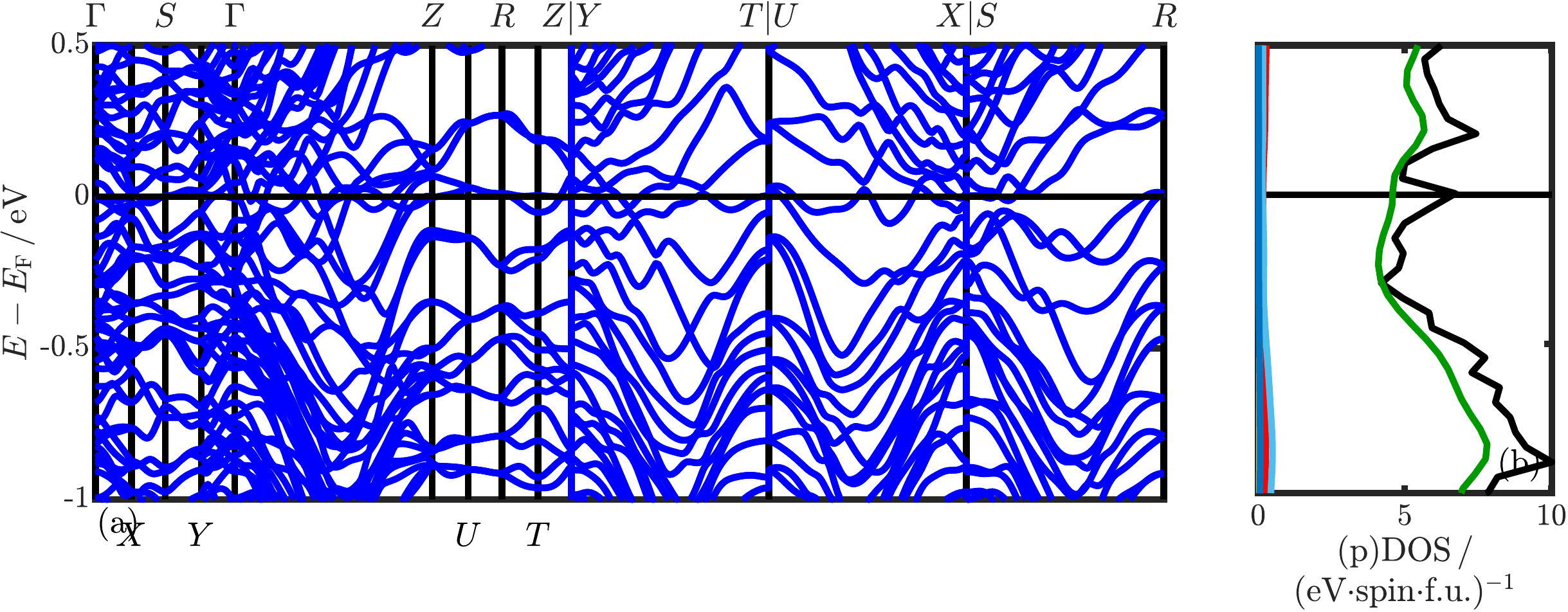}
	\caption{
		\ce{Nb14S5} at \SI{0}{\giga\pascal} 
	} 
	\label{fig:Nb14S5_0}
\end{figure}
 
\begin{figure}[ht]
	\includegraphics[width=0.77\columnwidth]{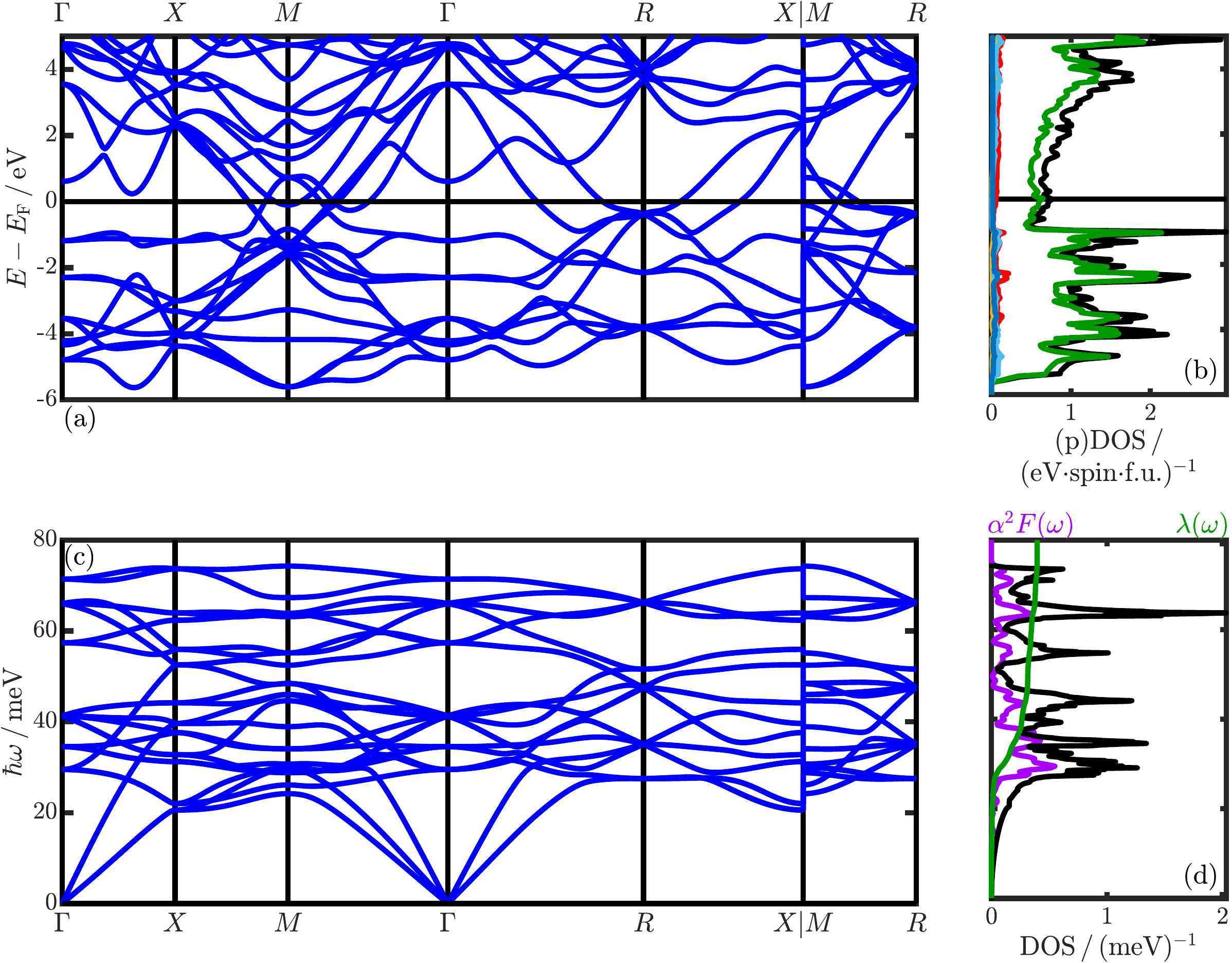}
	\caption{
		\ce{Nb3S} at \SI{200}{\giga\pascal}
	} 
	\label{fig:Nb3S_200}
\end{figure}

%\FloatBarrier
%\null
%\vfill
%\newpage
 
\begin{figure}[ht]
	\includegraphics[width=0.77\columnwidth]{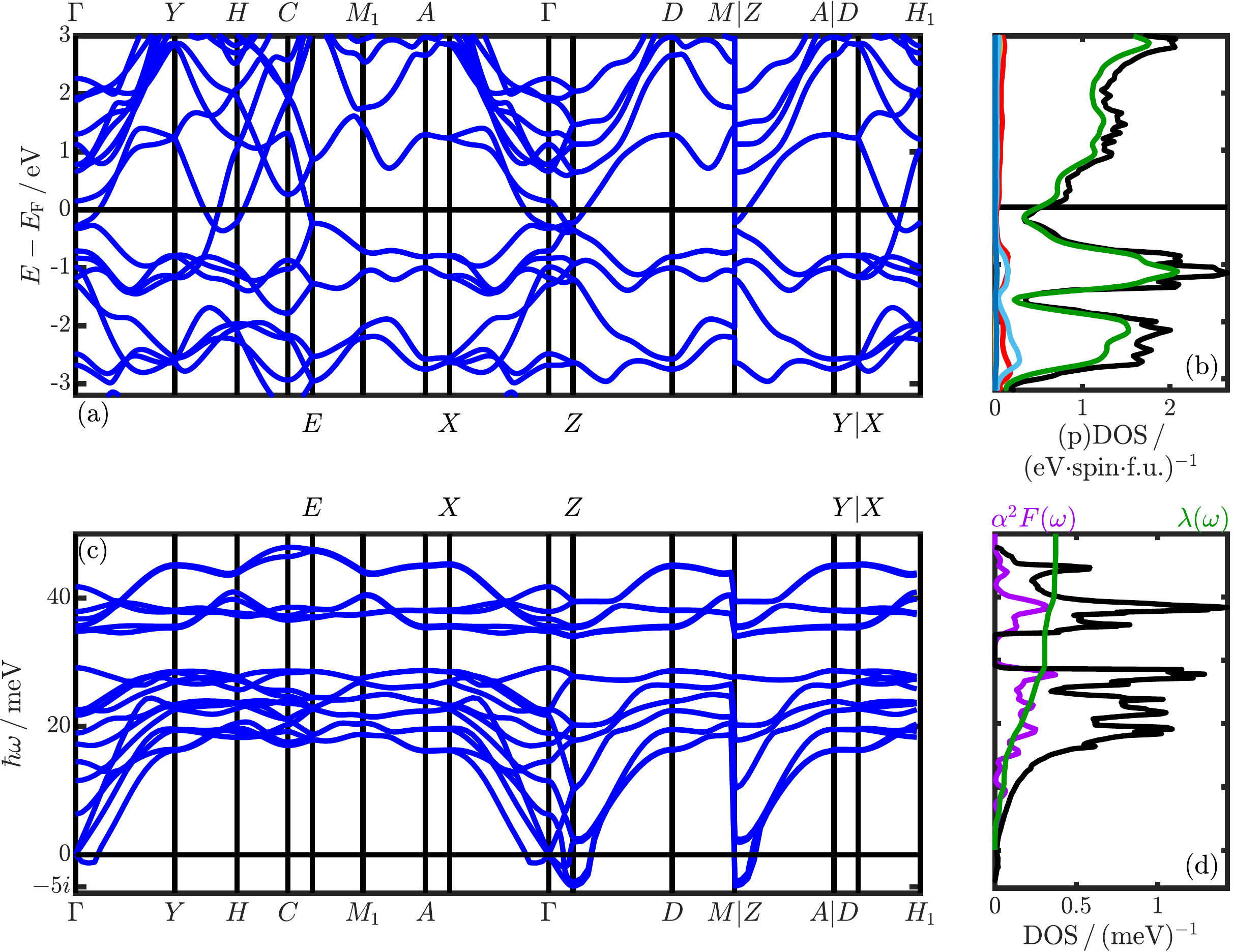}
	\caption{
		\ce{Nb2S} at \SI{0}{\giga\pascal} 
	} 
	\label{fig:Nb2S_0}
\end{figure}
 
\begin{figure}[ht]
	\includegraphics[width=0.77\columnwidth]{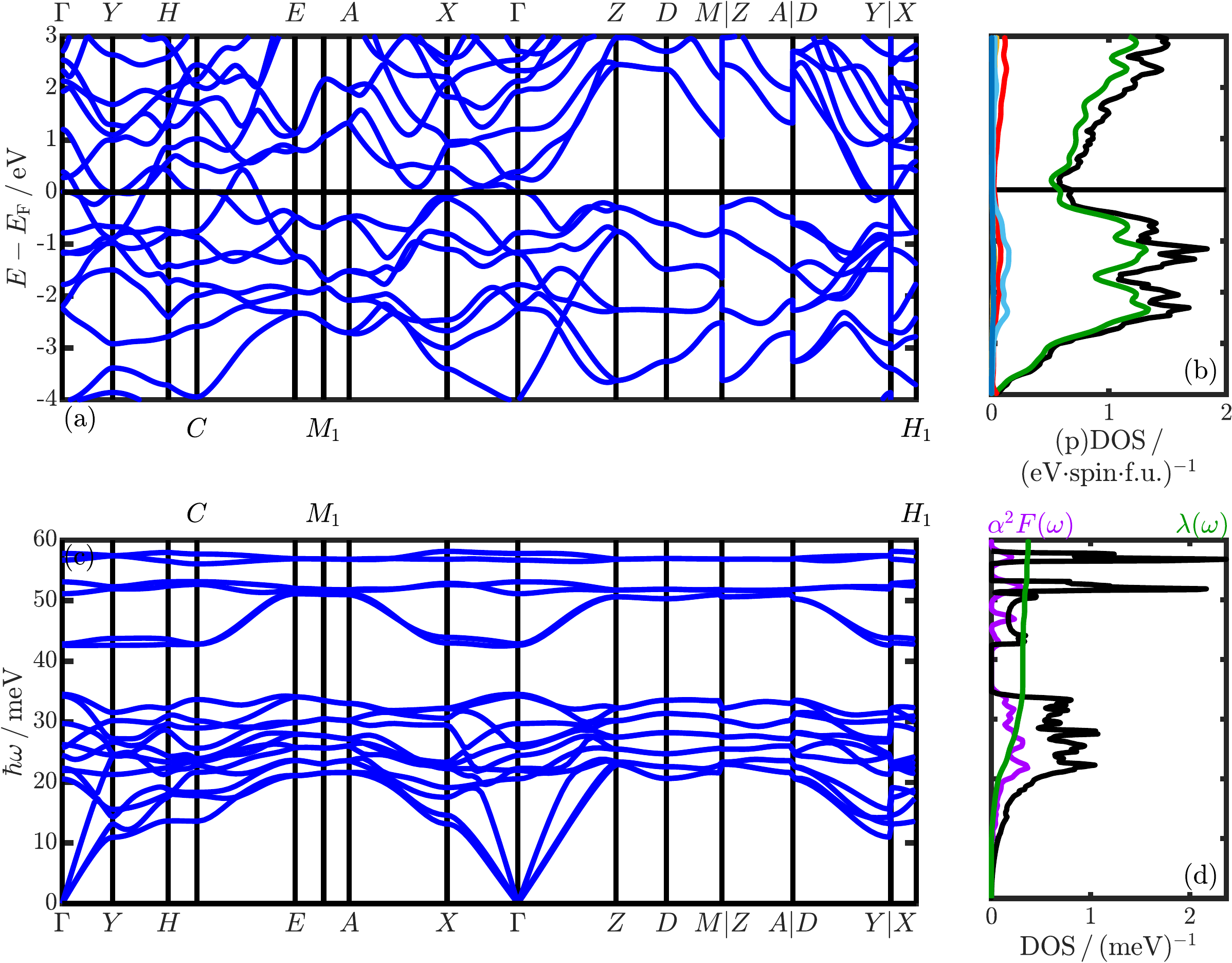}
	\caption{
		\ce{Nb2S} at \SI{25}{\giga\pascal}
	} 
	\label{fig:Nb2S_25}
\end{figure}
 
\begin{figure}[ht]
	\includegraphics[width=0.77\columnwidth]{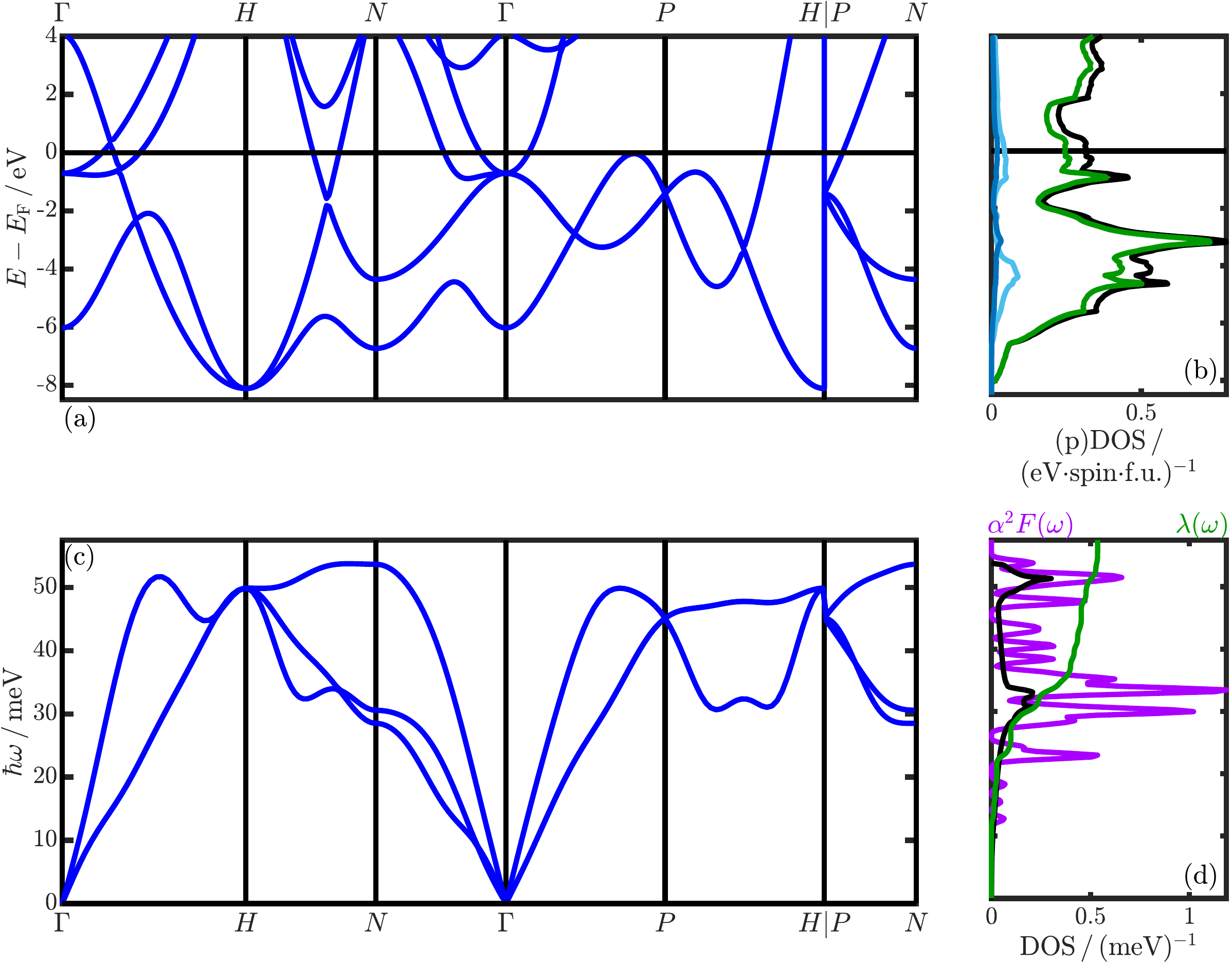}
	\caption{
		\ce{Nb} at \SI{200}{\giga\pascal}
	} 
	\label{fig:Nb_200}
\end{figure}

\begin{figure}[ht]
	\includegraphics[width=0.77\columnwidth]{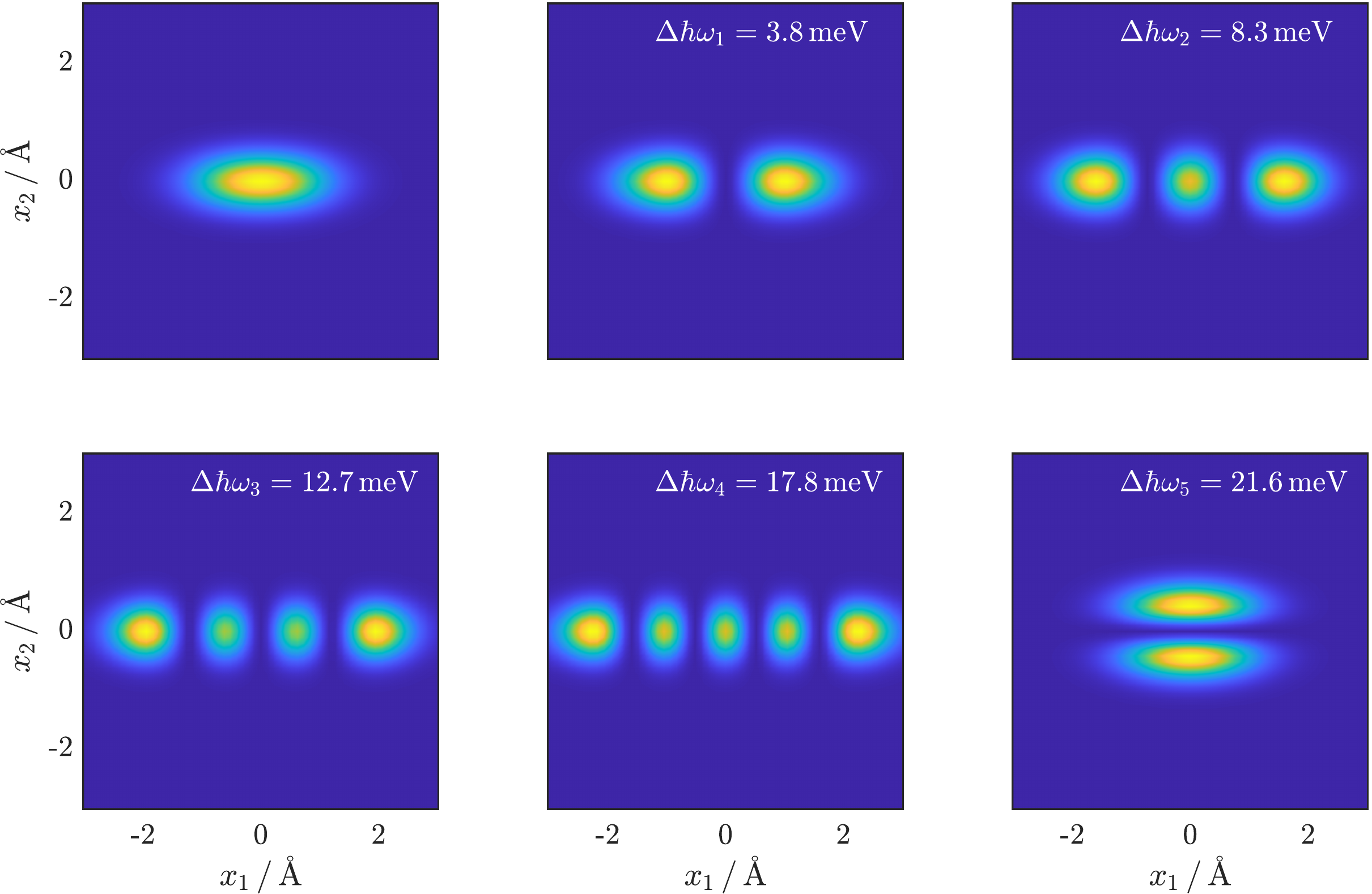}
	\caption{
		Square moduli of the first six wavefunctions, together with their eigenenergies (with respect to the lowest state) calculated for the 2-dimensional APES of Nb$_2$S at 225~GPa, taking into account the coupling between the first two phonon modes at $A$. The first four excited states belong to the first phonon mode and the fifth to the second phonon mode. The approximated eigenfrequencies for the two modes ($\Delta \hbar \omega_1$ and $\Delta \hbar \omega_5$) are in good agreement with the uncoupled, 1-dimensional APES calculations (3.8 and 21.6 meV vs. 3.1 and 20.4 meV, respectively, 1D solutions not shown explicitly).
	} 
	\label{fig:SEQ_2D_A}
\end{figure}

\begin{figure}[ht]
	\includegraphics[width=0.77\columnwidth]{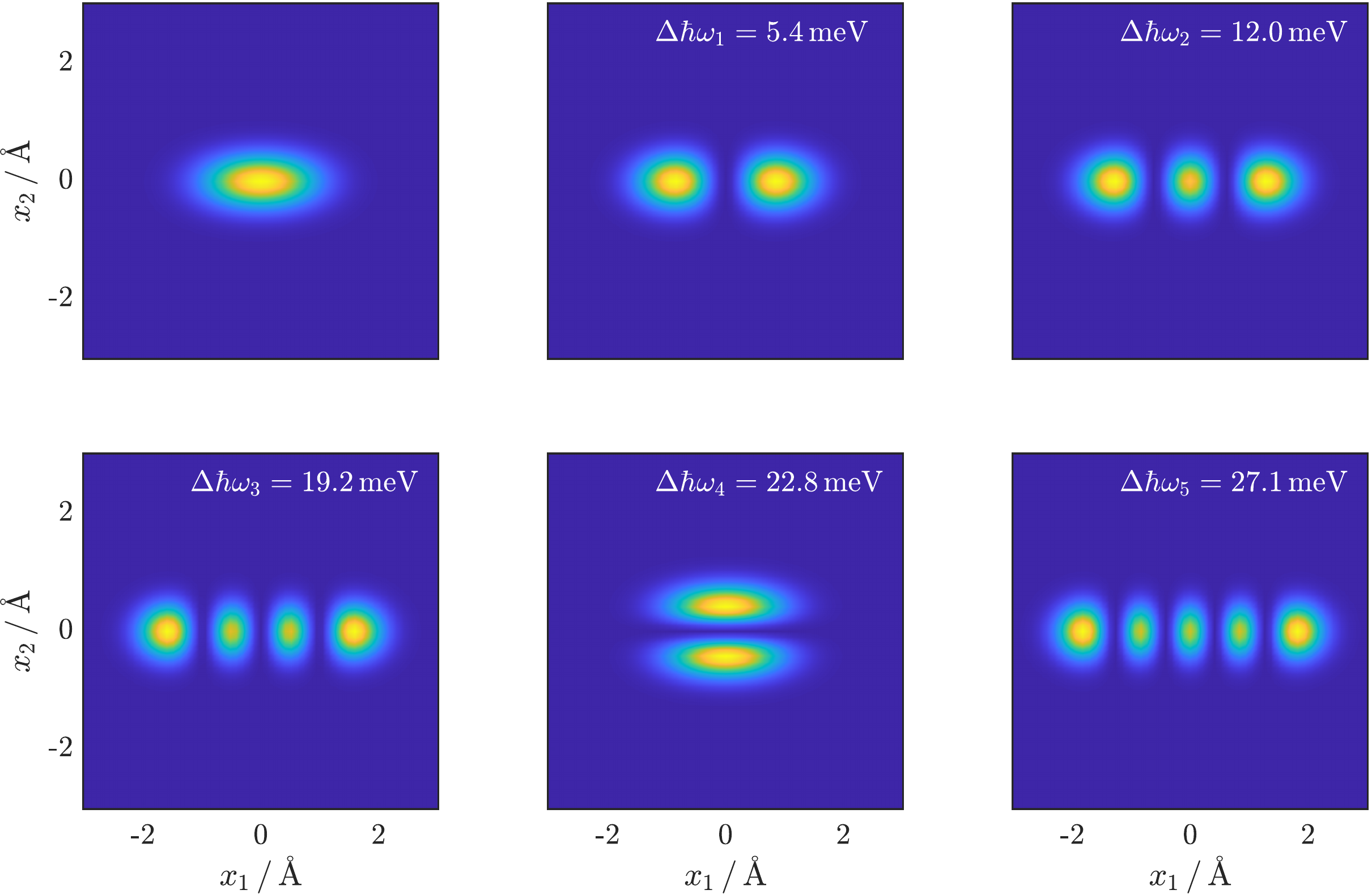}
	\caption{
		Square moduli of the first six wavefunctions, together with their eigenenergies (with respect to the lowest state) calculated for the 2-dimensional APES of Nb$_2$S at 225~GPa, taking into account the coupling between the first two phonon modes at $2/3 \, A$. The first three and the fifth excited states belong to the first phonon mode and the fourth to the second phonon mode. The approximated eigenfrequencies for the two modes ($\Delta \hbar \omega_1$ and $\Delta \hbar \omega_4$) are in good agreement with the uncoupled, 1-dimensional APES calculations (5.4 and 22.8 meV vs. 3.5 and 24.9 meV, respectively, 1D solutions not shown explicitly).
	} 
	\label{fig:SEQ_2D_23}
\end{figure}

\FloatBarrier
%\vfill
